\documentclass{aa}  

\usepackage{graphicx}
\usepackage{txfonts}
\usepackage{amssymb}
\usepackage{xcolor}
\usepackage{subcaption}
\usepackage{csquotes}

\usepackage{hyperref}

\newcommand{\hi}{H\,{\sc i}}
\newcommand{\co}{CO}
\newcommand{\hii}{H\,{\sc ii}}
\newcommand{\Htwo}{H\textsubscript{2}}
\newcommand{\Halpha}{\text{H}$\alpha$}
\newcommand{\ZAP}{\textsc{zap}}

\defcitealias{cardelliRelationshipInfraredOptical1989}{C89}
\defcitealias{emsellemPHANGSMUSESurveyProbing2022}{E22}
\defcitealias{calzettiDustContentOpacity2000}{C00}

\begin{document}
    \title{Tidal pre-conditioning and ram-pressure stripping in NGC\,1427A}
    \subtitle{Deep VLT/MUSE spectroscopy and FUV--to--radio observations trace a Fornax Cluster dwarf in transformation}

    \author{
        J. P. Carvajal\inst{1}\fnmsep\thanks{Corresponding author: Jcarvajal000@gmail.com}
            \and
        T. H. Puzia\inst{1}
            \and
        M. Bla\~na\inst{2}
            \and
        P. K. Nayak\inst{1}
            \and
        K. Fahrion\inst{3}
            \and
        M. Hilker\inst{4}
            \and
        E. Emsellem\inst{4}
            \and
        E. D. Skillman\inst{5}
            \and
        M. D. Mora\inst{6}
            \and
        B. Miller\inst{7}
            \and
        Y. Ordenes-Brice\~no\inst{8}
            \and
        P. Eigenthaler\inst{9, 10}
            \and
        R. Rahatgaonkar\inst{1}
            \and
        J. Chanam\'e\inst{1}
          }

    \institute{
    Institute of Astrophysics, Pontificia Universidad Católica de Chile, Av. Vicuña Mackenna 4860, 7820436 Macul, Santiago, Chile
        \and
    Centro Espacial Nacional, Fuerza Aérea de Chile, Av. Pedro Aguirre Cerda 5500, Cerrillos, Santiago, Chile
        \and
    Department of Astrophysics, University of Vienna, T\"{u}rkenschanzstra{\ss}e 17, 1180 Wien, Austria
        \and
    European Southern Observatory, Karl-Schwarzschild-Stra{\ss}e 2, 85748 Garching bei M\"unchen, Germany
        \and
    Minnesota Institute for Astrophysics, University of Minnesota, 116 Church
Street SE, Minneapolis, MN 55455, USA
        \and
    Las Campanas Observatory, Carnegie Observatories, Casilla 601, La Serena, 7820436, Chile
        \and
    Gemini Observatory, South Operations Center, Casilla 603, La Serena, Chile
        \and
    Instituto de Estudios Astrofísicos, Facultad de Ingeniería y Ciencias, Universidad Diego Portales, Av. Ejército Libertador 441, Santiago, Chile
        \and
    Instituto de Astrof\'isica, Universidad Andr\'es Bello, Fern\'andez Concha 700, 7591538 Las Condes, Santiago, Chile
        \and
    Max Planck Institute for Astronomy, K\"onigstuhl 17, 69117 Heidelberg, Germany
    }

    \date{First revision \today}
    \abstract
    {The early environmental transformation of low-mass cluster galaxies from gas-rich to gas-poor remains poorly constrained, in part because clear, phase-resolved observations are rare. NGC\,1427A, a disturbed star-forming dwarf in the Fornax cluster, offers a favorable case for studying this process.}
    {We aim to build a spatially resolved, multi-phase picture of NGC\,1427A in order to constrain the roles of ram-pressure stripping and tidal perturbations in its present transformation.}
    {We combine a deep, spatially contiguous VLT/MUSE mosaic with ancillary data from the FUV to the radio. Full-spectrum fitting of the MUSE cube yields maps of stellar kinematics, ages, metallicities, and continuum attenuation, while emission-line modeling provides ionized-gas kinematics, Balmer-decrement reddening, and star-formation-rate surface densities. The ancillary multi-wavelength data trace the neutral and molecular gas, dust, and recent star formation, placing the MUSE-based results in a broader multi-phase context.}
    {We find a pronounced decoupling between stars and gas: the \hi\ and ionized gas rotate about an axis tilted with respect to the stellar field and are globally blueshifted. Joint constraints from stellar and nebular attenuation, infrared dust tracers, and \hi\ morphology indicate stripping with a strong line-of-sight component that has progressed into the ISM. 
    At the same time, the asymmetric distribution of gas and dust, together with the structured and time-dependent star formation, points to an additional gravitational perturbation, with a recent mild fly-by by a nearby dwarf being the favored interpretation.}
    {We propose that dwarf--dwarf tidal effects have torqued and pre-conditioned the gas, while the Fornax intracluster medium is driving ram-pressure stripping that now reaches the ISM and coincides with a declining global star formation rate. This places NGC\,1427A at the onset of environmentally driven quenching, making it a useful benchmark for early cluster dwarf transformation.}
   \keywords{
    Galaxies: dwarf --
    Galaxies: individual: NGC 1427A --
    Galaxies: clusters: individual: Fornax --
    Galaxies: interactions --
    Galaxies: ISM --
    Galaxies: evolution
               }

   \maketitle

\section{Introduction}

Low-mass galaxies in clusters are heavily transformed by their environment, yet the relative roles of ram-pressure stripping (RPS) and gravitational tides, and, crucially, how the collisionless stellar component and collisional (ionized and neutral gas, and dust) remain coupled or decouple, are still debated \citep[e.g.][]{mayerSimultaneousRamPressure2006, corteseDawesReviewRole2021a, boselliRamPressureStripping2022a}. Compared to field dwarfs, present-day cluster dwarfs are systematically redder, more quiescent, and structurally transformed, reflecting accelerated quenching and morphological change in dense environments \citep[e.g.][]{boselliRamPressureStripping2022a, wangEvolutionaryContinuumNucleated2023, brownVERTICOVIIEnvironmental2023, chambaImpactEnvironmentSize2024, Blana2025}. However, the transitional path between field-like and cluster-processed dwarfs remains poorly constrained, complicating like-for-like comparisons of scaling relations across environments \citep[e.g.][]{mistaniAssemblyDwarfGalaxies2016,romero-gomezAreEarlytypeGalaxies2024}. Observationally, key uncertainties persist in understanding timescales, viewing geometry, and the disentanglement of hydrodynamical and tidal effects \citep[e.g.][]{poggiantiGASPGasStripping2017, boselliRamPressureStripping2022a,corteseDawesReviewRole2021a,poggiantiMUSEViewRam2025}. The scarcity of clear Ram Pressure Stripping (RPS) ``snapshots'' may reflect rapid processing: simulations suggest efficient removal of circumgalactic gas on $\sim$few$\,\times\,10^8$\,yr timescales, shortening the observable transition window in low-mass systems \citep[e.g.][]{benitez-llambayDwarfGalaxiesCosmic2013, zhuItsBreezeCircumgalactic2024, ghoshRamPressureStripping2024}. Furthermore, the accretion history leaves an imprint on the kinematics and morphology of stars and cold gas, and their misalignment, underscoring the importance of multi-phase tracers \citep[e.g.][]{zengKinematicMorphologyLowmass2024}. Spatially contiguous, phase-resolved observations are therefore essential to connect stellar structure and kinematics to the evolving gas and dust during environmental transformation.

The Fornax galaxy cluster is an ideal nearby laboratory for such studies. It is compact and dynamically evolved, at a distance $D\simeq19.3\pm0.7$\,Mpc \citep{anandTRGBSBFProjectTip2024, blakesleeACSFORNAXCLUSTER2009a}, with well-mapped substructure and internal dynamics \citep{drinkwaterSubstructureDynamicsFornax2001a, drinkwaterEvolutionStarFormation2001a, Schuberth10, chaturvediFornaxClusterVLT2022, reiprichSRGEROSITAAllsky2025}. Its total mass is comparatively modest \citep[$M_{\rm vir}\!\approx\!7\times10^{13}\,{\rm M_\odot}$][]{drinkwaterSubstructureDynamicsFornax2001a}, potentially extending the timescales over which early gas removal and transformation can be observed. Deep wide-field imaging from the Next Generation Fornax Survey (NGFS) and the Fornax Deep Survey (FDS) has enabled a detailed census of dwarfs and low-surface-brightness structure \citep[e.g.][]{munozUNVEILINGRICHSYSTEM2015a, eigenthalerNextGenerationFornax2018a, venholaFornaxDeepSurvey2018a, rajFornaxDeepSurvey2019a}. On the gas side, the MeerKAT Fornax Survey reveals a subpopulation of \hi-rich dwarfs, some with disturbances or tails \citep{serraMeerKATFornaxSurvey2023a, kleinerMeerKATFornaxSurvey2023a}, tracing gas loss. These assets, together with the distance-modulus precisions that can be achieved with JWST for 3-D placement \citep{anandTRGBSBFProjectTip2024}, make Fornax uniquely suited to dissect the roles of RPS and tides in dwarf galaxies in cluster environments.

\begin{figure*}
    \centering
    \includegraphics[width=\linewidth]{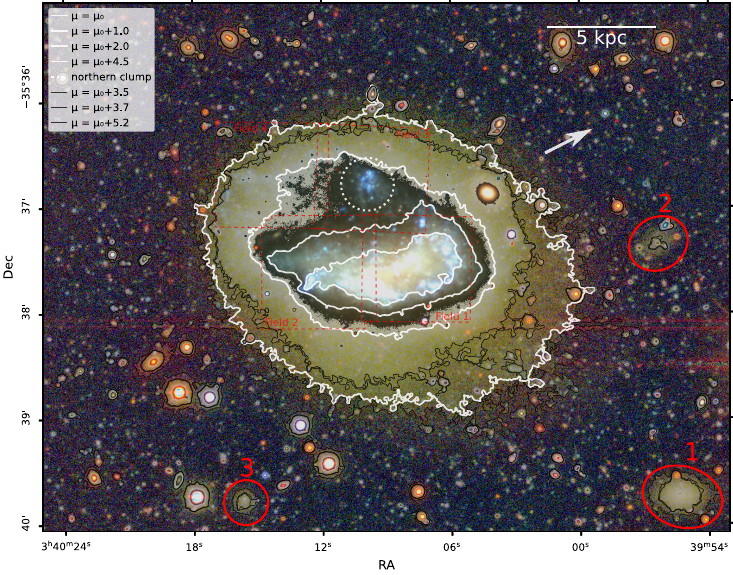}
    \caption{NGFS DECam $i^{\prime}g^{\prime}u^{\prime}$ (RGB) view of NGC\,1427A. North is up, east is left; the arrow marks the direction to the Fornax cluster center. The MUSE mosaic footprint is shown in transparent red. White and black curves trace $u^{\prime}{+}g^{\prime}{+}i^{\prime}$ isophotes at different surface-brightness levels (see legend). The dashed white circle marks the ``northern clump'' \citep{lee-waddellTidalOriginNGC1427A2018a}; labeled red ellipses mark nearby dwarf candidates discussed in the text and listed in Table~\ref{tab:nearby-dwarfs}.}
    \label{fig:ngc1427a-ngfs-isophotes}
\end{figure*}

NGC\,1427A is a star-forming Fornax dwarf with an arrowhead/cometary morphology and a one-sided \hi\ tail pointing in the anti-cluster-centric direction \citep[e.g.][]{hilkerNGC1427ALMC1997a, chanameIonizedGasKinematics2000, lee-waddellTidalOriginNGC1427A2018a, serraMeerKATFornaxSurvey2024}. Early work combining ATCA \hi\ with deep optical imaging suggested a tidal or merger origin for the disturbed stellar structure \citep{lee-waddellTidalOriginNGC1427A2018a}. A different interpretation was proposed by \citet{mastropietroTaleTwoTails2021}, who showed with simulations that the combined effect of the Fornax tidal field, galaxy rotation, orbital curvature, and RPS can reproduce a two-tail morphology resembling NGC\,1427A if the galaxy is only $\sim200~{\rm kpc}$ in front of the cluster center. A legacy globular cluster luminosity function distance places NGC\,1427A marginally in the foreground relative to the Fornax core, albeit with large uncertainties \citep[we adopt $D=17\pm1.5\text{ [stat.]}\pm1.0\text{ [syst.]}$\,Mpc throughout this work; see][]{georgievOldGlobularCluster2006}. Together with its large line-of-sight velocity offset relative to NGC\,1399 (about $600~\mathrm{km\,s^{-1}}$ larger), this supports a high-velocity infall scenario in which the cluster environment can play a central role \citep{chanameIonizedGasKinematics2000}. The deeper MeerKAT analysis of \citet{serraMeerKATFornaxSurvey2024}, however, revealed a more extended and kinematically complex \hi\ structure, and favored a scenario in which gas that was already tidally disturbed by a galaxy--galaxy encounter or merger was subsequently shaped by RPS. The complex optical low-surface-brightness environment, including an asymmetric outer envelope and nearby dwarf candidates, is illustrated in Fig.~\ref{fig:ngc1427a-ngfs-isophotes}.

Here we present results from a four-field VLT/MUSE mosaic of NGC\,1427A with a uniform reduction optimized for heterogeneous observing conditions (see Tab.~\ref{tab:observations}), providing contiguous coverage across the main body and outskirts (Fig.~\ref{fig:ngc1427a-ngfs-isophotes}). We analyze the MUSE mosaic alongside MeerKAT \hi\ maps \citep{kleinerMeerKATFornaxSurvey2023a}, deep NGFS imaging \citep{munozUNVEILINGRICHSYSTEM2015a}, AstroSat/UVIT FUV imaging, and Spitzer/Herschel infrared data \citep{sivanandamTracingRampressureStripping2014,fullerHerschelFornaxCluster2014}. These data enable a spatially registered, multi-phase view of the stars, ionized gas, neutral gas, and dust.

\begin{table}[!ht]
\caption{Summary of MUSE observations for NGC\,1427A.}
\label{tab:observations}
\centering
\begin{tabular}{l l c c c c}
\hline\hline
Prog. ID & Field & PI & Sky$^{a}$ & Exp.\ (s) & Mode$^{b}$\\
\hline
094.B$-$0612 & (1) & Mora    & -              & 21$\times$967                      & no \\
110.23VM & (2)     & Fahrion & \(\checkmark\) & 27$\times$540                      & AO \\
110.23VM & (3)     & Fahrion & \(\checkmark\) & 20$\times$540$^{c}$                & AO \\
111.24R5 & (4)     & Mora    & -              & 3$\times$967$^{c}$                 & no \\
\hline
\end{tabular}
\tablefoot{
\tablefoottext{a}{Offset-sky frames available (\(\checkmark\)) or not (-).}
\tablefoottext{b}{Instrument mode: ‘AO' = WFM-AO, ‘no' = WFM-noAO.}
\tablefoottext{c}{A subset of exposures was rejected from the final coadds owing to quality defects.}
}
\end{table}

\section{Data}
\label{sec:data}

\subsection{MUSE observations and data reduction}
\label{sec:data_muse}

The VLT/MUSE observations were obtained between 2015 and 2023, using WFM-noAO and WFM-AO mode (Tab.~\ref{tab:observations}). The data were reduced with the ESO MUSE pipeline \citep[v3.13.8;][]{weilbacherDataProcessingPipeline2020}, complemented by custom procedures designed for the galaxy's extended low-surface-brightness outskirts and heterogeneous observing conditions.

At a high level, the workflow comprises (i) standard calibrations and construction of science pixel tables, (ii) astrometric alignment of each exposure to a common reference frame using an archival \textit{HST}/ACS image (Prog. ID 9689), (iii) sky modeling and subtraction with field-specific strategies, and (iv) coaddition onto a common output WCS to form the final cube. Fields~2 and~3 include dedicated offset-sky exposures, Field~4 contains sufficient blank sky for in-field modeling, while Field~1 is entirely filled by galaxy emission and therefore uses a sky-continuum estimate anchored to the overlap with Fields~2--3. Residual sky features are mitigated using \ZAP\ \citep{sotoZAPEnhancedPCA2016}. After this procedure, the flux levels between fields agree to within $\lesssim$10\% in pairwise overlap regions. We also construct a no-\ZAP\ control mosaic to verify that \ZAP\ does not introduce artifacts in either emission-line measurements or the continuum.

A detailed description of the reduction steps is provided in Appendix~\ref{app:MUSEdatared}.

\subsection{MUSE data analysis}
\label{sec:data_analysis}

We derive high-level products from the final MUSE mosaic (Sect.~\ref{sec:data_muse}; App.~\ref{app:MUSEdatared}) using a workflow similar in spirit to widely used data-analysis pipelines (DAPs) for resolved optical spectroscopy, such as TIMER \citep{gadottiTIMER-MUSE2019}, MaNGA \citep{belfioreDataAnalysisPipeline2019,westfallDataAnalysisPipeline2019}, and PHANGS-MUSE \citep[][E22 hereafter]{emsellemPHANGSMUSESurveyProbing2022}. These approaches typically build on full-spectrum fitting and simultaneous line modeling with tools such as \textsc{pPXF} \citep{cappellariParametricRecoveryLineofSightPPXF2004,cappellariImprovingFullSpectrumPPXF2017}. We treat the wavelength-dependent MUSE line-spread function (LSF) as described in \citepalias{emsellemPHANGSMUSESurveyProbing2022}; key configuration choices and derived estimators are summarized in App.~\ref{app:DAPmethods}.

We analyze the \ZAP-cleaned cube with a modified version of \textsc{tardis}\footnote{\url{https://gitlab.com/francbelf/ifu-pipeline}}, the open-source implementation of the PHANGS-MUSE DAP \citepalias{emsellemPHANGSMUSESurveyProbing2022}, which itself builds on \textsc{gist}\footnote{\url{https://gitlab.com/abittner/gist-development}} \citep{bittnerGISTPipelineMultipurpose2019}. We adopt \textsc{tardis} because it is extensively tested on star-forming systems \citepalias{emsellemPHANGSMUSESurveyProbing2022} and, crucially for a disturbed dwarf with spatially varying star formation \citep[e.g.][]{mcquinnTrueDurationsStarbursts2009}, its default stellar-population fits do not impose regularization. Throughout, we construct maps directly from the \textsc{pPXF} outputs to retain explicit control over masking and quality cuts, and to avoid propagating measurements into bins with poor fits (see App.~\ref{app:DAPmethods}).

\paragraph{Preparation and masking.}
We adopt a reference systemic velocity of $2035$~km~s$^{-1}$ \citep[e.g.][]{tonrySBFSurveyGalaxy2001,serraMeerKATFornaxSurvey2024} and a Milky-Way foreground reddening $E(B{-}V)=0.01$ \citep{schlaflyMeasuringReddeningSloan2011}. Foreground stars and unrelated background sources are masked prior to binning and fitting; these sources are not analyzed further. We apply adaptive Voronoi binning \citep{cappellariAdaptiveSpatialBinning2003} to reach a uniform continuum S/N across the mosaic, computing S/N in the 5300--5500\,\AA\ window from the propagated variance. We adopt a baseline target $\mathrm{S/N}=35$ and repeat the full analysis at $\mathrm{S/N}=60$ and $90$ as a robustness check; key maps and integrated quantities are consistent, with the expected trade-off between spatial resolution and statistical noise.

\paragraph{Stellar continuum: kinematics, attenuation, and populations.}
We fit the stellar continuum with \textsc{pPXF} using E-MILES Simple Stellar Population (SSP) models \citep{vazdekisUVextendedEMILESStellar2016} assuming a \citet{chabrierGalacticStellarSubstellar2003} IMF and BaSTI isochrones \citep{pietrinferniLargeStellarEvolution2004}. We follow the \citetalias{emsellemPHANGSMUSESurveyProbing2022} fitting sequence,  but use extended low-metallicity SSP grids better suited to dwarf galaxies; the exact kinematic and stellar-population template subsets are listed in App.~\ref{app:DAPmethods}. The sequence consists of (i) a stellar-kinematics fit measuring the first four Gauss-Hermite moments ($V$, $\sigma$, $h_3$, $h_4$), (ii) a dedicated attenuation step to constrain $E(B{-}V)_*$ while mitigating template/continuum degeneracies \citep[cf.][]{sextlTYPHOONStellarPopulation2024,sextlTYPHOONStellarPopulation2025}, and (iii) a non-regularized stellar-population fit over a broad MUSE wavelength range to recover template weights and their uncertainties via Monte Carlo realizations \citepalias{emsellemPHANGSMUSESurveyProbing2022}. Strong night-sky residual regions are masked during the fits (Fig.~\ref{fig:DQ-Overview}; gray bands show the masked spectral windows used in the fits). In the attenuation step, we disable both additive and multiplicative Legendre polynomials; the polynomial scheme in the other \textsc{pPXF} runs follows \citetalias{emsellemPHANGSMUSESurveyProbing2022} and is summarized in App.~\ref{app:DAPmethods}.

For internal attenuation within NGC\,1427A, we adopt a Calzetti-type extinction curve with an LMC-like total-to-selective ratio $R_V=3.41$ \citep{calzettiDustContentOpacity2000, gordonQuantitativeComparisonSmall2003}, and apply this prescription consistently to both the stellar continuum and nebular emission.

\paragraph{Emission-line modeling.}
Nebular emission lines are fitted with \textsc{pPXF} after subtracting the best-fitting stellar continuum, tying kinematics within three line families (hydrogen recombination, low-ionization, and high-ionization species) following \citet{belfioreDataAnalysisPipeline2019}. Line amplitudes are fit freely, with fixed branching ratios where required by atomic physics (e.g., common strong doublets); we adopt a single kinematic component per family. To improve stability in the red, we mask the brightest sky lines in the observed frame using the UVES sky-emission atlas \citep{hanuschikUVESSkyLineAtlas2003} as a reference. We fit emission lines at the Voronoi-bin level and at the spaxel level initialized from the $\mathrm{S/N}=35$ solution. Lines are considered robust where they are not fully compromised by masking and reach $\mathrm{S/N}>5$ in at least one bin.

\paragraph{Derived maps.}
From the fitted products we construct maps of stellar kinematics, stellar attenuation, and stellar-population summaries (e.g., surface-mass density and mass-weighted mean age/metallicity), together with emission-line fluxes and gas kinematics. Where needed, we convert extinction-corrected \Halpha\ luminosities into SFRs using standard calibrations \citep{kennicuttStarFormationMilky2012,chomiukUnificationStarFormation2011} and Balmer-decrement reddening under case-B assumptions \citep[with a Calzetti-type curve;][C00 hereafter]{calzettiDustContentOpacity2000}. The exact estimators, quality cuts, and uncertainty propagation used for derived quantities (including Balmer-decrement reddening and \Halpha-based SFRs) are summarized in Appendix~\ref{app:DAPmethods}.

\subsection{Ancillary data}
\label{sec:data_ancillary}

To place the MUSE results in a multi-phase context, we combine the optical ionized-gas and stellar diagnostics with ancillary constraints on the neutral and molecular gas, recent star formation, and dust. In practice, we rely on (i) MeerKAT \hi\ products from the MeerKAT Fornax Survey \citep{serraMeerKATFornaxSurvey2023a, kleinerMeerKATFornaxSurvey2023a, serraMeerKATFornaxSurvey2024}, which provide \hi\ column density and kinematics at arcsecond-to-kiloparsec scales in the Fornax environment, (ii) a deep UVIT FUV image in the F148W band \citep{rampazzoDoradoItsMember2022, tandonInorbitCalibrationsUltraviolet2017, tandonAdditionalCalibrationUltraviolet2020}, used as an independent tracer of recent ($\lesssim 100$ Myr) star formation alongside \Halpha, (iii) infrared constraints from \textit{Spitzer}/IRAC imaging and \textit{Spitzer}/IRS spectroscopy together with \textit{Herschel} far-infrared photometry \citep{ordenes-bricenoNGFS-IV-IR2018, smithSpectralMappingReconstruction2007, fullerHerschelFornaxCluster2014}, which trace PAH-sensitive emission and the warm/cold dust components, and (iv) the ALMA \co(1--0) non-detection reported by \citet{zabelALMAFornaxCluster2019a}, which bounds the CO-bright molecular gas reservoir.

For the qualitative multiwavelength comparisons presented in this work, we register ancillary maps to the MUSE astrometric frame; where appropriate (e.g., UV/IR overlays), we homogenize the effective resolution by PSF-matching to a common Gaussian FWHM set by the lowest-resolution dataset in the comparison. We restrict the main text to the aspects of these data products that directly support the interpretation of the MUSE results, while technical details (UVIT calibration, IR processing choices, and derived global quantities such as dust SED fits and CO-to-\Htwo\ limits) are reported in Appendix~\ref{app:ancillary}.

\section{Results}
\label{sec:results}

\begin{table}
\caption{Nearby dwarf galaxies within $\sim10\arcmin$ of NGC\,1427A.}
\label{tab:nearby-dwarfs}
\centering
\begin{tabular}{l l c}
\hline\hline
\# & Denomination & $\log(M_*/M_{\odot})^{a}$ \\
\hline
1 & FCC\,229$^{b}$              & 7.6 \\
2 & WFLSB\,1$-$6$^{c}$          & 7.1 \\
3 & NGFS\,034016$-$353946$^{d}$ & 6.9 \\
\hline
\end{tabular}
\tablefoot{
\tablefoottext{a}{Stellar masses from the NGFS.}
\tablefoottext{b}{\citet{fergusonPopulationStudiesGroups1989}; alternate denominations NGFS\,033955$-$353943 and FDS11\_DWARF134.}
\tablefoottext{c}{\citet{mieskeEarlytypeDwarfGalaxy2007a}; alternate denomination NGFS\,033956$-$353721 \citep{eigenthalerNextGenerationFornax2018a}.}
\tablefoottext{d}{\citet{eigenthalerNextGenerationFornax2018a}; alternate denomination FDS11\_DWARF140. NGFS and FDS dwarf catalogs from \citet{eigenthalerNextGenerationFornax2018a} and \citet{venholaFornaxDeepSurvey2018a}, respectively.}
}
\end{table}

\subsection{Multi-phase morphology and environment}\label{subsec:results-morph-and-context}
\begin{figure}
    \centering
    \includegraphics[width=\linewidth]{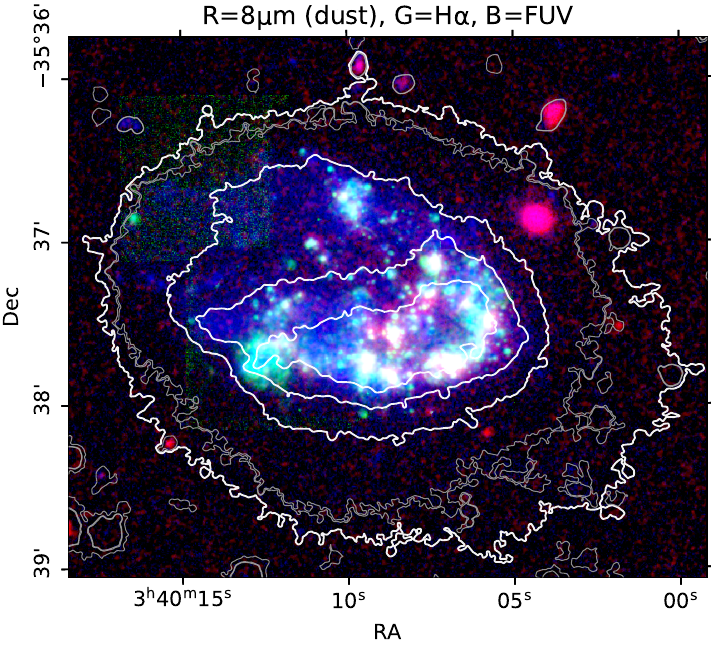}
    \caption{Three-color composite highlighting distinct star formation and ISM tracers in NGC\,1427A: IRAC 8~$\mu$m (red; stellar-continuum subtracted), MUSE \Halpha\ (green), and UVIT FUV (blue). All images are PSF-matched to a common 2\arcsec\ FWHM Gaussian. In the main body, 8~$\mu$m emission envelopes \Halpha\ knots as expected for dusty \hii\ regions; the northern clump shows clear 8~$\mu$m emission, with faint bridge-like 8~$\mu$m connecting to the main body. By contrast, several high-surface-brightness \Halpha\ clouds at the leading edge (toward the Fornax cluster center) exhibit little or no 8~$\mu$m, indicating ionized gas with weak co-spatial aromatic/small-grain emission. We also note extended FUV emission to the northeast with weak \Halpha\ counterparts, consistent with $\sim$50-100~Myr populations after ionizing stars have faded. See Section~\ref{sec:data} and Appendix~\ref{app:ancillary} for details.}
    \label{fig:FUV-HA-8UM}
\end{figure}

The NGFS image in Figure~\ref{fig:ngc1427a-ngfs-isophotes} shows that NGC\,1427A has a characteristic arrowhead/cometary appearance with bright, clumpy star-forming knots embedded in an irregular low-surface-brightness (LSB) envelope. At $\mu\simeq\mu_0+3.5$ the isophotes depart from simple ellipses with a southward extension, and the faintest levels ($\mu\simeq\mu_0+5$) are elongated toward the southwest. A very faint, stream-like feature is tentatively traced in that direction. Several LSB dwarfs within a projected radius of $\sim10\arcmin$ ($\sim60$~kpc) are listed in Table~\ref{tab:nearby-dwarfs} with literature identifiers and stellar-mass estimates. The most massive of those, FCC\,229, lies approximately along the axis of the faint southwest extensions. None shows an obvious substructure at our depth, with only FCC\,229 having a nuclear star cluster offset by 1\arcsec from the center (see App.~\ref{app:optical}). We defer membership and dynamical relevance to the discussion, but note here that their projected configuration naturally allows for fast fly-by scenarios. Toward the northeast, the LSB envelope appears noticeably bluer, suggesting a younger halo population, consistent with the measured FUV emission (Fig.~\ref{fig:FUV-HA-8UM}).

\begin{figure}
    \centering
    \includegraphics[width=\linewidth]{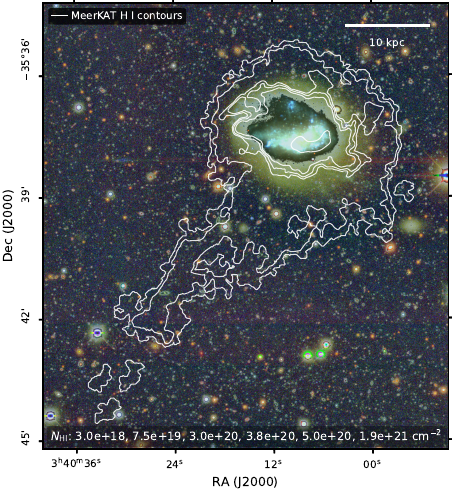}
    \caption{NGFS $u^\prime g^\prime i^\prime$ RGB image of NGC~1427A with MeerKAT \hi\ column-density contours (6$^{\prime\prime}$ beam) overlaid in white. Contours are drawn at several column densities, tracing the concentration of \hi, the gaseous disk and the southwest tail around the stellar body.}
    \label{fig:rgb_meerkat_contours}
\end{figure}

The optical morphology captured by NGFS is put into context by the MeerKAT \hi\ column density map. Figure~\ref{fig:rgb_meerkat_contours} overlays \hi\ contours at levels
$N_{\mathrm{H\,\textsc{i}}}=3.0\times10^{18}\times5^{\,n}\ \mathrm{cm^{-2}}$
(increasing integer $n$) on the NGFS color image. Three features stand out:
(i) two tails, one in the anti-cluster-centric direction (southeast) and another pointing toward FCC\,229 (see also \citealt{loniBlindATCAHI2021a,mastropietroTaleTwoTails2021,serraMeerKATFornaxSurvey2024}),
(ii) an elongation from northeast to southwest at $N_{\mathrm{H\,\textsc{i}}}=3.8\times10^{20} \mathrm{cm^{-2}}$, and
(iii) the highest column densities offset to the southwest relative to the optical body.
Taken together, these patterns show that the gas distribution is disturbed on multiple spatial scales, with one component aligned roughly anti-cluster-centrically and another along the northeast--southwest axis. We return to the possible origin of these components, including RPS, cluster tides, and galaxy--galaxy interactions, in Sect.~\ref{sec:discussion}.

\begin{figure*}[!ht]
    \centering
    \includegraphics[width=\textwidth]{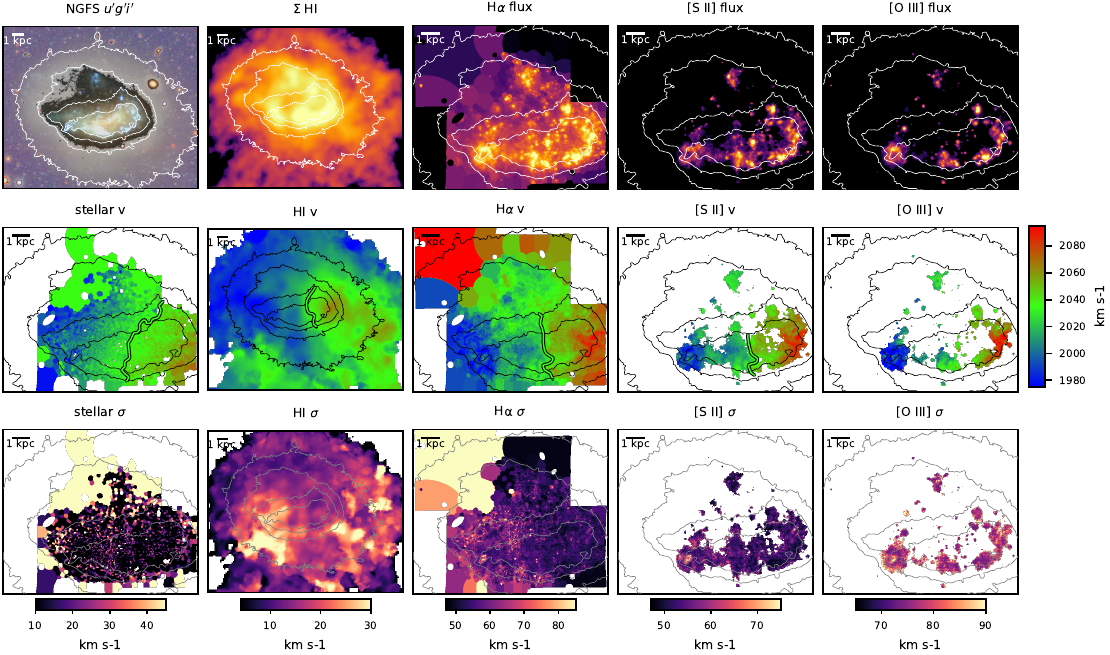}
    \caption{
    \textbf{Multi-phase overview of NGC\,1427A.}
    Each column shows the same tracer; each row shows a different property. The maps for MUSE-based emission lines show only measurements with ${\rm S/N}>$10 (i.e., relative flux error $<10\%$). For visualization, ionized-gas velocity and dispersion maps are hybrid: spaxel-level values where ${\rm S/N}>10$, Voronoi-bin values elsewhere.
    \textit{Row\,1 - surface brightness}: 
    NGFS $u'g'i'$ composite, MeerKAT H\,\textsc{i} column density ($6''$ beam), and MUSE maps of \Halpha, [S\,\textsc{ii}]$\,\lambda\lambda 6716,6731$, and [O\,\textsc{iii}]$\,\lambda\lambda 4959,5007$ flux.
    \textit{Row\,2 - line-of-sight velocity}: stellar pPXF velocity, H\,\textsc{i} moment--1, and hybrid ionized--gas velocity maps (spaxel-level where ${\rm S/N}>10$, Voronoi-bin values elsewhere). All five panels share the color bar on the right. In all but the last panel, the systemic-velocity contour ($v=2035$\,km\,s$^{-1}$) is highlighted within the $\mu_0+2$ isophote to guide the eye to the approximate zero-velocity line.
    \textit{Row\,3 - velocity dispersion}: 
    stellar, H\,\textsc{i}, \Halpha, [S\,\textsc{ii}], and [O\,\textsc{iii}] dispersions with individual color bars below each panel.
    Overlaid on each frame are the contours for spatial reference. These isophotes were calculated over the NGFS $u'+g'+i'$ image. From the brightest isophote with surface brightness $\mu_0$, the others show $\mu_0+1$, $\mu_0+2$, and $\mu_0+4.5$, with no spatial smoothing. A 1\,kpc bar is shown in the top-left corner of every panel (distance adopted: 17\,Mpc).
    }
    \label{fig:ngc1427a_15panel}
\end{figure*}

The displacement of the bulk \hi\ is similarly evidenced by other tracers of collisional phases, such as ionized gas via optical emission lines (Fig.~\ref{fig:ngc1427a_15panel}), and warm dust via \textit{Spitzer} spectral mapping (Fig.~\ref{fig:MIR-FIR-dust}). The spatial distribution of the non-collisional component is best shown by the pPXF MUSE-inferred stellar surface density map (see Fig.~\ref{fig:sfh-trip-mcavg}; see also 3.6~$\mu$m imaging in \citet{sivanandamTracingRampressureStripping2014}). These observations evidence decoupling between the collisional (gas and dust) and non-collisional (stars) phases.

\subsection{Multi-phase kinematics}

Figure~\ref{fig:ngc1427a_15panel} summarizes the stellar and gaseous kinematics together with surface-brightness tracers. MeerKAT H\,\textsc{i} maps trace the cold atomic phase, MUSE constrains the stars and the gas via Balmer and ionized element emission lines.

A global feature in the velocity maps is an offset between the stellar and gaseous kinematics. To quantify this in a homogeneous way, we estimate the dominant large-scale velocity gradient of the stellar, H$\alpha$, and \hi\ velocity fields within a common elliptical aperture tracing the main stellar body, and define the corresponding kinematic major-axis position angle from the direction of that gradient (see Appendix~\ref{app:kinpa} for details). With position angles measured east of north, we obtain ${\rm PA}_{\rm kin}=63.6^\circ \pm 1.4^\circ$ for the stellar component, ${\rm PA}_{\rm kin}=91.4^\circ \pm 0.5^\circ$ for H$\alpha$, and ${\rm PA}_{\rm kin}=102.0^\circ \pm 1.8^\circ$ for \hi. These values confirm a clear kinematic misalignment of order $\sim30^\circ$ between the stellar and gaseous components, while H$\alpha$ and \hi\ remain more nearly aligned with each other.

\begin{figure}
    \centering
    \includegraphics[width=0.9\linewidth]{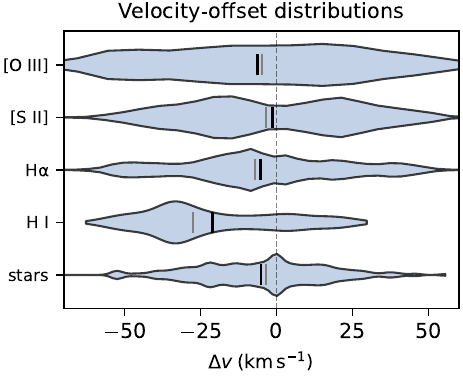}
    \caption{Violin plot of the line-of-sight velocity residuals of individual components with respect to the systemic radial velocity: $\Delta v\!=\!v-2035\,\mathrm{km\,s}^{-1}$, where 2035\,km\,s$^{-1}$ is our adopted systemic velocity. From top to bottom: residuals in [O\,\textsc{iii}] (high-ionization), [S\,\textsc{ii}] (low-ionization), \Halpha\ (Balmer emission), H\,\textsc{i}, and the stellar component, measured inside the $\mu_0+2.0$ isophote. Each violin shows the kernel-density estimate (vertical width $\propto$ probability), while the gray line marks the median, the black line the mean, and the dashed vertical line highlights $\Delta v=0$. Distributions are computed from the velocity maps in Fig.~\ref{fig:ngc1427a_15panel}.}
    \label{fig:deltaV_violinplot}
\end{figure}

Within the main body (inside the $\mu_0$ and $\mu_0+1$ isophotes), the \Halpha\ and \hi\ surface-brightness distributions and velocities broadly coincide. Northward, \hi\ becomes increasingly blueshifted relative to the ionized clumps. Velocity residuals inside the $\mu_0+2$ isophote are defined as $\Delta v \equiv v - 2035~\mathrm{km\,s^{-1}}$ (our adopted systemic radial velocity; negative values denote blueshift). As shown in Figure~\ref{fig:deltaV_violinplot}, the \hi\ residuals are skewed to negative values, with a mean offset of $\sim-16~\mathrm{km\,s^{-1}}$ and a median of $\sim-24~\mathrm{km\,s^{-1}}$ relative to the stars. \Halpha\ residuals are closer to the stellar distribution but are marginally skewed to negative values, with a median of $-3.5~\mathrm{km\,s^{-1}}$. Low- and high-ionization tracers ([S\,\textsc{ii}], [O\,\textsc{iii}]) have incomplete coverage within this isophote. However, where measured, their kinematics are consistent with \Halpha. The \hi\ map has a beam size of $\gtrsim6''$. Beam smearing broadens the velocity distribution and may dilute peak offsets.

\subsection{Stellar-population properties}
\label{sec:stellar_pop_properties}

\begin{figure*}
    \centering
    \includegraphics[width=\textwidth]{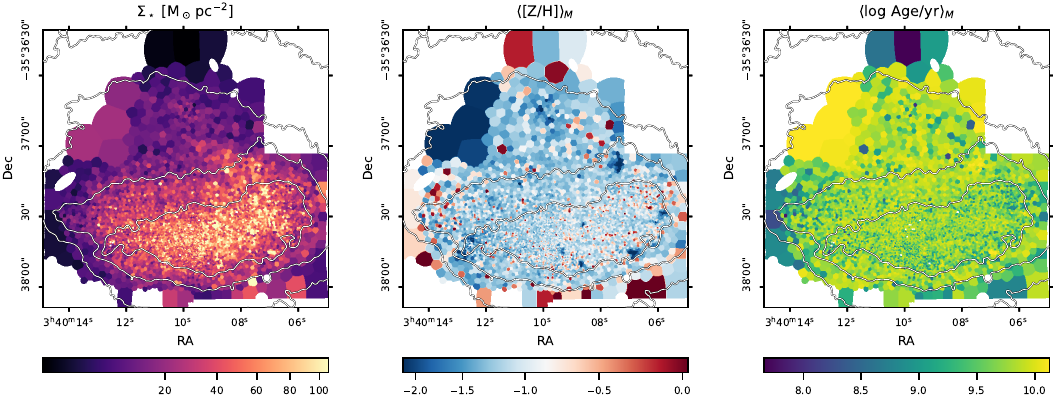}
    \caption{From left to right, we show the stellar mass surface density ($\Sigma_*$) averaged across MC realizations, the mass-weighted metallicity ($\langle\mathcal{Z}\rangle_{\rm M}$), and the mass-weighted age ($\langle\mathcal{A}\rangle_{\rm M}$). Only bins with continuum S/N $>30$ are shown, close to the target of 35.}
    \label{fig:sfh-trip-mcavg}
\end{figure*}

\begin{figure*}
    \centering
    \includegraphics[width=1.0\textwidth]{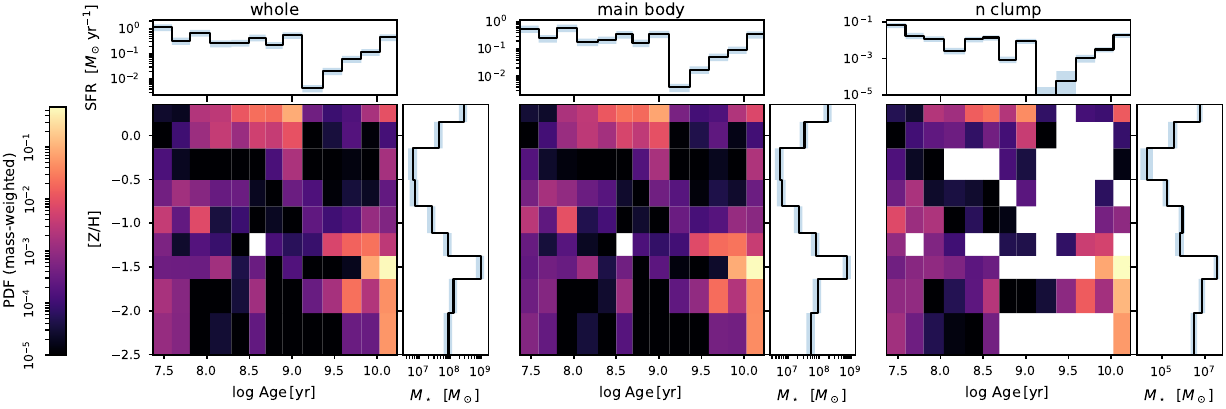}
    \caption{
    Region-integrated stellar age--metallicity distributions and their 1-D projections.
    Each column corresponds to a spatial region (whole galaxy, main body, northern clump).
    The heatmap shows $\mathrm{PDF}^{\mathrm{reg}}$ on the native SSP $(\log t,[Z/H])$ grid, defined as the region-summed \textsc{pPXF} template masses normalized to unity (i.e., the fraction of the region's present-day stellar mass assigned to each template; Appendix~\ref{app:pdfreg}).
    Top panels: star-formation histories derived by converting the mass-weighted distribution to formed mass using the SSP living-mass fraction, marginalizing over metallicity, and dividing by the linear time-bin width to yield $\mathrm{SFR}(t)$.
    Side panels: metallicity distribution functions obtained by marginalizing over age and scaling by the region's present-day stellar mass.
    }
    \label{fig:stellar-pdfs-apertures}
\end{figure*}
We map the stellar mass surface density, $\Sigma_*$, the mass-weighted mean metallicity, $\langle \mathcal{Z} \rangle_M$, and the mass-weighted mean logarithmic age, $\langle \mathcal{A} \rangle_M \equiv \langle \log_{10}(\mathrm{age}/\mathrm{yr}) \rangle_M$ shown in Figure~\ref{fig:sfh-trip-mcavg} (see also Sect.~2.2 and App.~B.2). For each bin, these quantities are derived from the pPXF SSP template masses returned over the Monte Carlo ensemble: $\Sigma_\ast$ is the total present-day stellar mass in the bin divided by its projected area, while $\langle \mathcal{Z} \rangle_M$ and $\langle \mathcal{A} \rangle_M$ are the corresponding mass-weighted means over the SSP metallicity and age grid. Figure~\ref{fig:stellar-pdfs-apertures} presents the region-integrated age--metallicity distributions and their one-dimensional projections for the whole galaxy, the main body, and the northern clump.

To summarize the stellar-population fits over a region, we combine the pPXF template masses across all Voronoi bins in that region and normalize them so that the total sums to unity. The resulting 2D distribution on the SSP age--metallicity grid, therefore, represents the fraction of present-day stellar mass assigned to each (age, [Z/H]) template. One-dimensional star-formation histories are obtained by converting present-day mass to initially formed mass using the SSP living-mass fractions and dividing by the age-bin width; metallicity distribution functions are obtained by marginalizing over age and converted into present-day stellar mass.

The stellar surface-mass density peaks across the main body, coincident with the solid-body-like stellar rotation, and declines smoothly toward the outskirts. Optical light peaks over-emphasize regions with recent star formation relative to $\Sigma_*$, and the mass map illustrates the underlying stellar population structure. The northern clump does not appear strongly as a stellar overdensity. Its contrast is dominated by a bright, young component and local dust geometry, consistent with its prominence in \Halpha\ and FUV.

The region-integrated stellar age--metallicity distributions (Fig.~\ref{fig:stellar-pdfs-apertures}) show three prominent components: (i) a dominant old, metal-poor population in mass, (ii) a young (0.1-1.0~Gyr), metal-rich component that contributes strongly to the light, and (iii) very young ($\lesssim 0.1$~Gyr), metal-poor pockets. In the maps, some of the most metal-poor zones coincide with \Halpha-bright regions (Fig.~\ref{fig:sfh-trip-mcavg}; see also \citealt{Fahrion2026a}).

The combination of the \Halpha- and FUV-based star formation tracers, together with the pPXF-inferred stellar populations, provides complementary constraints on the star formation history from Myr to Gyr timescales (see \citealt{Fahrion2026a} for insights from star clusters). The youngest stellar-population templates in our grid have an age of 30~Myr. The recovered SFR in this youngest age bin ($\lesssim 40$~Myr) is $1.2^{+0.3}_{-0.4}~M_\odot\,\mathrm{yr^{-1}}$. For comparison, the FUV-based SFR is $0.117 \pm 0.041~M_\odot\,\mathrm{yr^{-1}}$ and the \Halpha-based SFR is $0.069 \pm 0.020~M_\odot\,\mathrm{yr^{-1}}$  (Table~\ref{tab:globalprops}). The relative offsets between these indicators and with respect to the star-formation main sequence (SFMS) are discussed in Sec.~\ref{sec:disc-sfr-sfh}.

The northern clump is marginally more metal-poor than the main body by $\sim 0.2$\,dex. Both the light- and mass-weighted ages are similar across apertures (Table~\ref{tab:globalprops}), indicating no marked differences in the SFH. Within the apertures, the main body contains $\sim 74\%$ of the stellar mass and the northern clump $\sim 3.4\%$.

\begin{table*}
\caption{Integrated properties of NGC\,1427A in three apertures.}
\label{tab:globalprops}
\centering
\begin{tabular}{lcccccc}
\hline\hline
 & \multicolumn{2}{c}{Whole galaxy}
 & \multicolumn{2}{c}{Main body}
 & \multicolumn{2}{c}{Northern clump}\\
Property & mean $\pm$ unc. & $\sigma_{\mathrm{int}}$
         & mean $\pm$ unc. & $\sigma_{\mathrm{int}}$
         & mean $\pm$ unc. & $\sigma_{\mathrm{int}}$\\
\hline
$\log_{10}(M_*/M_\odot)$
        & $9.28\pm0.13$ & \ldots
        & $9.10\pm0.13$ & \ldots
        & $7.76\pm0.13$ & \ldots\\[2pt]

SFR$_{{\rm H}\alpha}$\,$[10^{-3}\,M_\odot\,\mathrm{yr}^{-1}]$
        & $69\pm20$  & \ldots
        & $61\pm18$  & \ldots
        & $3\pm1$    & \ldots\\[2pt]

SFR$_{\rm FUV}$\,$[10^{-3}\,M_\odot\,\mathrm{yr}^{-1}]$
        & $117\pm41$ & \ldots
        & $94\pm33$  & \ldots
        & $8\pm3$    & \ldots\\[6pt]

$\langle\mathrm{[Z/H]}\rangle_{\rm MW}$
        & $-1.16\pm0.01$ & $0.75$
        & $-1.17\pm0.01$ & $0.72$
        & $-1.35\pm0.02$ & $0.63$\\[2pt]

$\langle\log_{10}\mathrm{Age/yr}\rangle_{\rm MW}$
        & $9.75\pm0.01$ & $0.64$
        & $9.76\pm0.01$ & $0.63$
        & $9.87\pm0.01$ & $0.60$\\[4pt]

$\langle\mathrm{[Z/H]}\rangle_{\rm LW}$
        & $-0.72\pm0.01$ & $0.83$
        & $-0.70\pm0.01$ & $0.80$
        & $-0.93\pm0.02$ & $0.79$\\[2pt]

$\langle\log_{10}\mathrm{Age/yr}\rangle_{\rm LW}$
        & $8.78\pm0.02$ & $0.95$
        & $8.82\pm0.02$ & $0.93$
        & $8.70\pm0.04$ & $1.08$\\[6pt]

$\log_{10}(M_{\rm HI}/M_\odot)$\,$^{\mathrm{a}}$
        & $9.22^{\mathrm{e}}\pm0.13$ & \ldots
        & \ldots & \ldots
        & \ldots & \ldots\\[2pt]

$\log_{10}(M_{\rm H_2,CO}/M_\odot)$\,$^{\mathrm{b}}$
        & $<\,7.28^{\mathrm{e}}$ & \ldots
        & \ldots & \ldots
        & \ldots & \ldots\\[2pt]

$\log_{10}(M_{\rm H_2,warm}/M_\odot)$\,$^{\mathrm{c}}$
        & $>7.4^{\mathrm{e}}$ & \ldots
        & \ldots & \ldots
        & \ldots & \ldots\\[6pt]

$\log_{10}(M_{\rm d,cold}/M_\odot)$\,$^{\mathrm{d}}$
        & $6.60\pm0.27$ & \ldots
        & \ldots & \ldots
        & \ldots & \ldots\\[2pt]

$\log_{10}(M_{\rm d,warm}/M_\odot)$\,$^{\mathrm{d}}$
        & $2.62\pm0.23$ & \ldots
        & \ldots & \ldots
        & \ldots & \ldots\\[1pt]

\hline
\end{tabular}
\tablefoot{
Quoted uncertainties on $\langle\mathrm{[Z/H]}\rangle$ and $\langle\log_{10}\mathrm{Age/yr}\rangle$ are formal errors propagated as described in the text; $\sigma_{\mathrm{int}}$ gives the intrinsic dispersion of the corresponding PDF where applicable. Errors on masses and SFRs that depend on distance are dominated by the adopted distance $D=17\pm1.5\text{ (stat.)}\pm1.0\text{ (syst.)}$\,Mpc.
\tablefoottext{a}{MeerKAT H\,\textsc{i} mass from \citet{serraMeerKATFornaxSurvey2024}.}
\tablefoottext{b}{3$\sigma$ upper limit on the molecular H$_2$ mass from
ALMA CO(1--0) observations \citep[][App.~\ref{app:radio_mm}]{zabelALMAFornaxCluster2019a}, rescaled
to our adopted distance.}
\tablefoottext{c}{Warm H$_2$ mass lower limit from the \textit{Spitzer}/IRS analysis of \citet[][]{sivanandamTracingRampressureStripping2014}, assuming OPR of 1.5.}
\tablefoottext{d}{Dust masses from the two-component modified-blackbody fit to the IRS LL + Herschel SED (App.~\ref{app:ir}), measured within an aperture comparable to the $\mu_0{+}1$ isophote. The total dust mass may be larger by up to a factor of $\sim2$ if the cold component follows the more extended 20--38\,$\mu$m emission out to $\mu_0{+}1$.}
\tablefoottext{e}{Values rescaled to the adopted distance of $17\pm2.5$\,Mpc; quoted uncertainties include distance contributions.}
}
\end{table*}

\subsection{Dust attenuation from stars and gas}
\label{sec:results-dust}

\begin{figure*}
    \centering
    \includegraphics[width=0.7\textwidth]{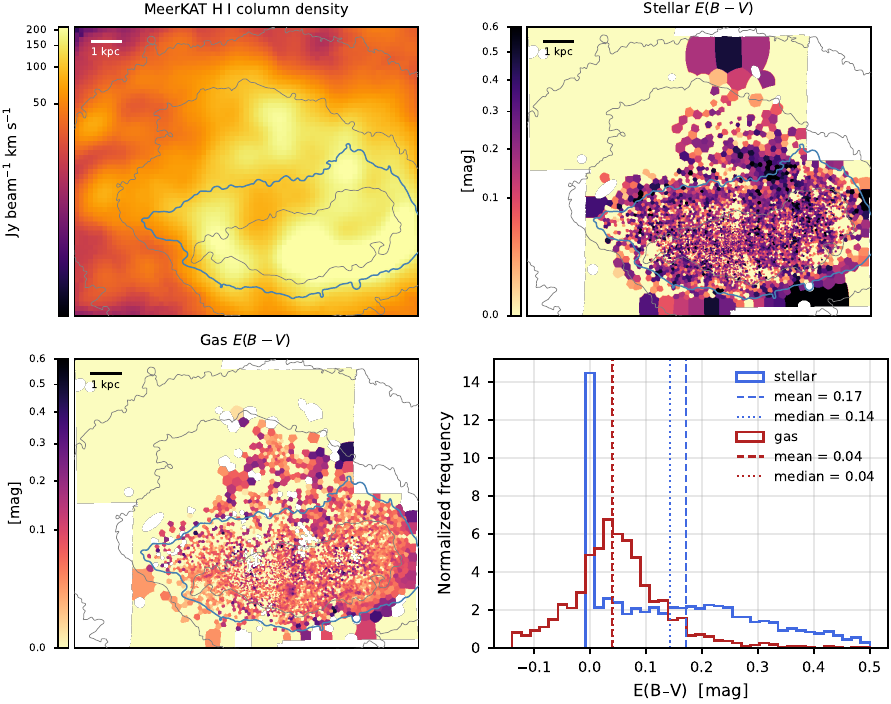}
    \caption{\textbf{Dust and gas context.} \textit{Top-left:} MeerKAT H\,\textsc{i} column density ($6''$ beam). \textit{Top-right:} stellar $E(B{-}V)_*$ from stellar continuum fit (App.~\ref{app:reddening_sfr}). \textit{Bottom-left:} nebular $E(B{-}V)_{\rm gas}$ from the Balmer decrement, masking bins with $\Delta E(B{-}V)_{\rm gas}>0.1$~mag. Both $E(B{-}V)$ maps share the same color scale. \textit{Bottom-right:} normalized histograms inside the $\mu_0{+}1$ isophote (blue: stars; red: gas); vertical lines mark the mean (dashed) and median (dotted). Contours show NGFS isophotes; a 1~kpc bar is shown in each map.}
    \label{fig:dust-properties}
\end{figure*}

We map stellar and nebular reddening using the estimators described in Section~\ref{sec:data_analysis} and Appendix~\ref{app:DAPmethods}. Figure~\ref{fig:dust-properties} shows (i) the MeerKAT \hi\ column density, (ii) stellar $E(B{-}V)_*$ from the pPXF stellar continuum fit, and (iii) nebular $E(B{-}V)_{\rm gas}$ from the Balmer decrement, together with (iv) their distributions inside the $\mu_0{+}1$ isophote. Approximately $90\%$ of the \Halpha\ emission originates from within this contour, and it contains most of the \hi\ gas. It is, therefore, the most representative region for the nebular emission and provides the fairest comparison with the stellar attenuation.

Inside the $\mu_0{+}1$ isophote, the stellar and nebular reddening distributions differ and we measure $\langle E(B{-}V)_*\rangle \simeq 0.17$ with median $\tilde E(B{-}V)_* \simeq 0.14$, versus $\langle E(B{-}V)_{\rm gas}\rangle \simeq 0.04$ with median $\tilde E(B{-}V)_{\rm gas} \simeq 0.04$.\footnote{Values from the Voronoi solutions using the LMC-like attenuation law as described in Section~\ref{sec:data_analysis}.} The stellar reddening positively correlates with the \hi\ column density, except in the brightest \Halpha\ regions where $E(B{-}V)_*$ is reduced and the continuum is dominated by young stellar populations. A highly obscured bridge between the main body and the northern clump coincides with $8\,\mu$m dust emission (see Fig.~\ref{fig:FUV-HA-8UM}, and Fig.~13 of \citealt{sivanandamTracingRampressureStripping2014}). \textit{Herschel} FIR images lack the resolution to trace this feature. We therefore use \hi\ as our primary tracer of cold, collisional matter. The implications of these dust constraints are discussed in Section~\ref{sec:rps-dust-geometry}.

\subsection{Current star-formation rate from \Halpha\ and FUV}
\label{sec:sfr_results}

\begin{figure}
    \centering
    \includegraphics[width=\linewidth]{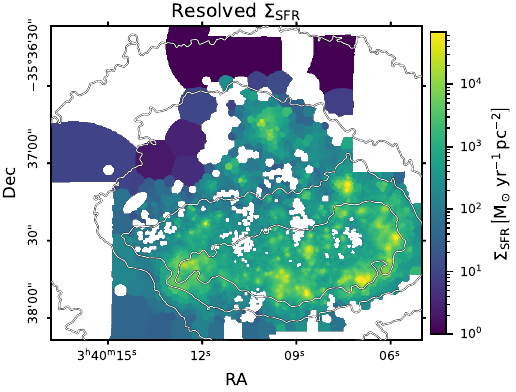}
    \caption{\textbf{Extinction-corrected \Halpha\ SFR surface density.}
    $\Sigma_{\rm SFR}$ computed per Voronoi bin using the Balmer-decrement-corrected \Halpha\ luminosity (App.~\ref{app:reddening_sfr}). Bins failing the S/N criteria ($\lesssim10$) or uncertainty cuts are masked (white).}
    \label{fig:resolved_sfr_map}
\end{figure}

Figure~\ref{fig:resolved_sfr_map} shows the \Halpha-based star-formation rate surface density ($\Sigma_{\rm SFR}$) map after Balmer-decrement correction (App.~\ref{app:DAPmethods}). Star formation is clumpy and widespread across the main body, with additional activity in the outer halo. The global SFR$_{\rm H\alpha}$ within the FoV is $\dot{M}_*\! =\! 0.069~M_\odot\,\mathrm{yr^{-1}}$ (Tab.~\ref{tab:globalprops}), while the main body contributes $\sim\!0.061~M_\odot\,\mathrm{yr^{-1}}$ and the northern clump contributes only $\sim\!0.003~M_\odot\,\mathrm{yr^{-1}}$.

The UVIT FUV data (Fig.~\ref{fig:FUV-HA-8UM}) shows a similarly patchy distribution, with a prominent northeast extension that is weak in \Halpha. Using the attenuation-corrected UVIT/F148W magnitudes and the $L_\nu$ calibration for a Kroupa/Chabrier IMF described in Appendix~\ref{app:uvit}, we obtain a global ${\rm SFR}_{\rm FUV} = 0.117 \pm 0.041~M_\odot\,\mathrm{yr^{-1}}$. To emphasize the influence of dust corrections, the corresponding bandpass-weighted attenuation is $A_{\rm FUV} = 0.492 \pm 0.201$\,mag for all three apertures. We assumed that the young stellar populations that dominate the FUV emission are subject to a similar dust geometry as the nebular emitting gas ($\langle E(B{-}V)_{\rm gas}\rangle \simeq 0.04$), but if the geometry is instead closer to the older stellar population, then FUV attenuation and star formation rates could be underestimated. Per-aperture FUV SFRs are reported in Table~\ref{tab:globalprops}.

On integrated scales, ${\rm SFR}_{\rm FUV}$ is modestly higher than ${\rm SFR}_{\rm H\alpha}$ (globally and for the main body, within uncertainties) and is comparatively enhanced in the northern clump (Table~\ref{tab:globalprops}). For the FUV attenuation correction, we adopt a single internal color excess consistent with the spatially averaged Balmer decrement reddening, implicitly assuming that the FUV-emitting populations experience similar dust columns to the nebular gas. If the FUV continuum is instead attenuated more like the older stellar population, then the inferred $A_{\rm FUV}$ and ${\rm SFR}_{\rm FUV}$ could be biased low. The statistical error budget is dominated by uncertainties in attenuation and distance; distance contributions can be neglected when comparing tracers internally within the galaxy. We discuss the physical interpretation of the offsets between tracers and their different characteristic timescales in Section~\ref{sec:disc-sfr-sfh}.

\section{Discussion}
\label{sec:discussion}

\subsection{The big picture of NGC 1427A's evolution}
\label{sec:big-picture}

NGC\,1427A is a low-mass, gas-rich dwarf that is currently undergoing strong environmental processing in the Fornax cluster. Throughout this work, we identify two main classes of environmental agents. The first is ram-pressure stripping (RPS) by the Fornax ICM, which has a significant global effect on the collisional, gaseous components. The long anti-cluster-centric \hi\ tail and the additional \hi\ extension toward the southwest \citep[][]{loniBlindATCAHI2021a, serraMeerKATFornaxSurvey2023a, serraMeerKATFornaxSurvey2024}, the systematic blueshift of \hi\ with respect to the stars, and the displacement of dusty gas along the line of sight all point to gas and dust being removed from the galaxy and pushed toward the observer. 
The unusual ``reddening inversion'' between stars and ionized gas (Sects.~\ref{sec:results-dust}~\&~\ref{sec:rps-dust-geometry}), together with the lack of obvious jellyfish-like \Halpha\ tentacles in projection \citep{Ebeling14, Poggianti16}, suggests that a substantial fraction of the dusty ISM associated with the inner disk has been displaced out of the disk plane and now lies in front of much of the stellar body along the line of sight.
The enhanced gas velocity dispersion seen in the southeastern part of the disk (Fig.~\ref{fig:ngc1427a_15panel}) is also consistent with turbulence and shear where gas is being displaced from the rotating disk by the ICM wind.

The second agent is a gravitational perturbation. RPS alone does not naturally explain the disturbance of the stellar body, the gas--star kinematic misalignment, and the off-center concentration of dense gas, dust, and recent star formation. Non-collisional tracers (i.e., stellar mass distribution and stellar kinematics; Fig.~\ref{fig:ngc1427a_15panel}) reveal a significant misalignment between the stellar and gaseous kinematics, while collisional tracers (\hi, ionized gas, warm dust, and FUV emission) are systematically displaced relative to the old stellar body. We also observe an FUV-bright, \Halpha-faint extension to the northeast, consistent with a $\sim\!50\!-\!100$~Myr-old star-forming episode.

Several tidal channels could, in principle, contribute to this disturbance. \citet{mastropietroTaleTwoTails2021} showed that the combined action of the Fornax tidal field, galaxy rotation, orbital curvature, and RPS can reproduce a two-tail morphology resembling NGC\,1427A. This remains a possible alternative scenario. However, the deeper MeerKAT analysis of \citet[][see their Sec. 4.2]{serraMeerKATFornaxSurvey2024} revealed a longer (70~kpc-long tail with scattered clouds out to 300~kpc) and more kinematically complex \hi\ structure than was available in the earlier ATCA-based view ($\sim$25~kpc-long tail), and favored a scenario in which gas that had already been tidally disturbed by a galaxy--galaxy encounter or merger was subsequently shaped by RPS. In this sense, the newer \hi\ evidence, together with the expected weakness of cluster tides at the present projected radius for a system of NGC\,1427A's mass density \citep{serraMeerKATFornaxSurvey2024, Blana2025}, shifts the balance away from a dominant role for the global Fornax tidal field and toward a galaxy--galaxy tidal perturbation plus RPS.

Our MUSE data support this mixed interpretation and add a recent-timescale constraint. The FUV/\Halpha\ morphology, the southwestern displacement of dense gas and dust, and the gas--star kinematic offset suggest that a recent or ongoing local perturbation has affected the ISM. Among the nearby dwarfs, FCC\,229 provides the most natural projected anchor for this recent/local component, as it lies to the southwest near the end of the secondary \hi\ extension. We therefore favor a recent mild fly-by involving FCC\,229 as a plausible contributor to the current off-center star formation and gas reorientation, while not requiring it to explain the full older large-scale stellar and \hi\ disturbance on its own.

Although NGC\,1427A lies close to the Fornax center in projection ($\sim$23\arcmin from NGC\,1399, i.e. $\sim$130~kpc at the Fornax distance), its true 3D cluster-centric distance remains uncertain. The legacy GCLF-based distance places it in the foreground of the Fornax core, albeit with large uncertainties. Nevertheless, the relevant point is that RPS can operate beyond the immediate cluster core in Fornax. eROSITA shows ICM emission and substructure to large radii (e.g. $R_{200}\approx604$~kpc), and long \hi\ tails are observed in Fornax galaxies at large projected separations along likely infall paths, such as NGC\,1437A at $\sim$73.6\arcmin, or $\sim$410~kpc, from NGC\,1399 (Fig.~14 of \citealt{reiprichSRGEROSITAAllsky2025}). 
Therefore, although the scenario of \citet{mastropietroTaleTwoTails2021}, in which NGC\,1427A lies $\sim$200~kpc in front of the cluster center and cluster tides play a major role, remains viable, it is not our preferred interpretation. The uncertain 3D distance prevents a purely geometrical discrimination between cluster tides and galaxy--galaxy perturbations. Our preference for a galaxy--galaxy tidal perturbation plus RPS is instead based mainly on the morphology and kinematics of the gas and stars, especially in light of the deeper MeerKAT analysis of \citet{serraMeerKATFornaxSurvey2024}. A quantitative orbit reconstruction remains beyond the scope of this work.

In the remainder of this section, we interpret our results within a mixed scenario in which RPS by the Fornax ICM acts on gas that had already been tidally disturbed. We first assess how far RPS has progressed through the different gas and dust phases, then examine the evidence for a localized recent perturbation, with FCC\,229 as the favored nearby candidate in projection. We then discuss how these processes have shaped the recent and past star formation history, the chemical and dust content of NGC\,1427A, and the properties of its surrounding substructures, before synthesizing the results into an evolutionary scenario.

\subsection{The extent of ram-pressure stripping}
\label{sec:extent-rps}

In this subsection, we examine how far ram-pressure stripping has progressed through the collisional components of NGC\,1427A: the neutral and ionized gas, and the dust. Taken together, these tracers indicate that ram pressure is not only removing gas from the circumgalactic regions but is also affecting the interstellar medium (ISM) of the main body, with a significant component along our line of sight.

\subsubsection{Neutral gas and kinematics}
\label{sec:rps-hi}

The \hi\ morphology and kinematics provide the clearest evidence for ongoing ram-pressure stripping. MeerKAT data reveal a long tail extending to the southeast, roughly anti-cluster-centric, as well as a secondary \hi\ enhancement toward the southwest (Fig.~\ref{fig:rgb_meerkat_contours}; \citealt{serraMeerKATFornaxSurvey2024}). Both features are consistent with gas being removed from the outer disk or circumgalactic medium (CGM) as NGC\,1427A moves through the Fornax ICM even at a large cluster-centric radius. In addition to these spatial asymmetries, the \hi\ is systematically blueshifted with respect to the stars by a median offset of $\sim-24$~km\,s$^{-1}$ (within the $\mu_0{+}2$ optical isophote; see Figs.~\ref{fig:ngc1427a-ngfs-isophotes}~\&~\ref{fig:deltaV_violinplot}), indicating that a substantial fraction of the neutral gas is being accelerated toward the observer. The line-of-sight velocity distribution (LOSVD) of the ionized gas is also skewed toward negative velocities, but the difference between its median velocity and that of the stars is only marginal. Larger LOSVD offsets between the velocities of different gas phases and the stars have been reported as signatures of ram-pressure stripping acting in other jellyfish systems \citep[e.g.][]{bellhouseGASPXV_LoSRPS_2019}.

These \hi-based signatures closely resemble those identified by \citet{serraMeerKATFornaxSurvey2024}, who used the U-shaped kinematical minor axis, the velocity gradient along the southern tail, and the lack of outer-disk compression to argue that NGC\,1427A is experiencing a predominantly line-of-sight, blueshifting ram-pressure wind, with the southeastern tail containing gas that was first displaced tidally and only later accelerated by ram pressure. Our analysis confirms and extends this picture with independent tracers (young and old stars, ionized gas, and dust) and allows us to quantify how far this line-of-sight ram pressure has propagated into the main body.

\begin{figure*}
    \centering
    \includegraphics[width=0.7\textwidth]{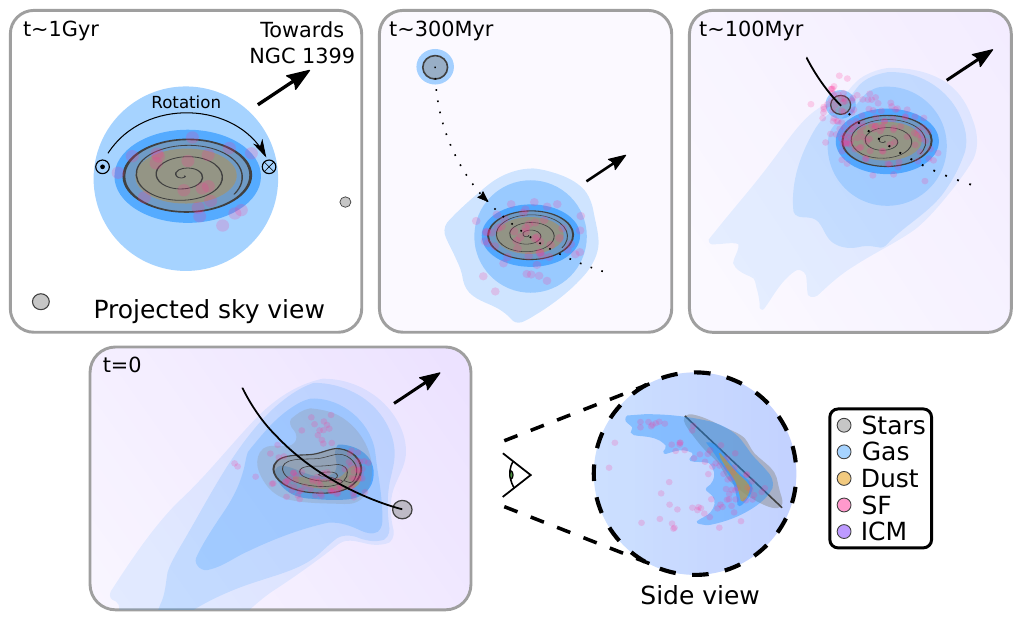}
    \caption{A cartoon of the likely evolutionary path described in this work.}
    \label{fig:scenario-cartoon}
\end{figure*}

The large projected extent of the long southeastern \hi\ tail, together with its substantial line-of-sight velocity gradient (from $\sim2035$~km\,s$^{-1}$ near the galaxy to $\sim1770$~km\,s$^{-1}$ in the farthest detached clouds along the tail; \citealt{serraMeerKATFornaxSurvey2024, Kleiner2025}), is naturally explained if NGC\,1427A follows a curved orbit through the Fornax potential. In that case, the material now forming the outer tail was likely removed more than 300~Myr ago \citep{serraMeerKATFornaxSurvey2024}, when the stripping direction had a stronger plane-of-sky component than it does today. The outer, mostly starless H\,I tail and the inner dusty ISM would then trace different stages of the same stripping event rather than the same instantaneous geometry. Figure~\ref{fig:scenario-cartoon} illustrates this interpretation, which we discuss further below. As an additional caution, a slight offset between the projected H\,I tail and the instantaneous anti-motion direction could partly reflect the imprint of gas-disk rotation on the stripped wake \citep[e.g.,][]{roedigerWakesRampressurestrippedDisc2006, roedigerRamPressureStripping2006}.

Beyond these global trends, the detailed \hi\ morphology may retain information about the three-dimensional disk geometry. 
Around the $N_\mathrm{H\,\textsc{i}}\sim4\times10^{20}$~cm$^{-2}$ contours in Fig.~\ref{fig:rgb_meerkat_contours}, the inner \hi\ distribution shows curved overdensities and extensions, giving the gaseous disk a disturbed and mildly twisted appearance, with weak spiral-arm-like features in a counterclockwise direction.
Observational and theoretical work shows that the vast majority of disk galaxies host trailing rather than leading spiral arms \citep[see][and references therein]{sellwoodSpiralsGalaxies2022}. This pattern suggests that the inferred \hi\ arms in NGC\,1427A are trailing. Together with the consistent velocity pattern seen across all tracers in Fig.~\ref{fig:ngc1427a_15panel} with the eastern side of the disk blueshifted and the western side redshifted, this favors a configuration in which the northern side of the disk is nearest to the observer, and the southern side is closer to the galaxy cluster. 
This orientation is also consistent with the dust-reddening geometry discussed in Sect.~\ref{sec:rps-dust-geometry}.

There are marked differences in the velocity-dispersion maps of \hi\ and \Halpha\ (bottom row in Fig.~\ref{fig:ngc1427a_15panel}). 
The western side of the galaxy appears relatively more turbulent in \hi, but exhibits a comparatively low \Halpha\ velocity dispersion. The eastern side, on the other hand, shows systematically higher velocity dispersion in both \hi\ and \Halpha. 
This is consistent with the scenario outlined earlier in which the neutral gas is being stripped toward the observer primarily on the western side, leaving behind the star-forming clumps and little velocity dispersion in the young stellar component, which implies that the neutral gas must have decoupled from the star-forming clumps $\gtrsim10$ Myr ago. 
The \Halpha\ velocity dispersion on the western side, tracing ionized gas associated with young star-forming regions, resembles the relatively featureless stellar velocity-dispersion map. The eastern side of the galaxy is more likely the result of fly-by-induced turbulence with more recent star-formation activity (See Sec.~\ref{sec:flyby_localised}).

\subsubsection{Geometry effects on dust attenuation}
\label{sec:rps-dust-geometry}

The dust distribution and its imprint on the stellar continuum and nebular emission confirm that ram pressure has started to affect the ISM within the main body. Most of the dust is cold and emits in the far-infrared, with a warmer component that is bright in the mid-infrared (Fig.~\ref{fig:MIR-FIR-dust}). \citet{fullerHerschelFornaxCluster2014} reported $\log(M_{\rm dust}/M_\odot)=6.62\pm0.12$ and $T_{\rm dust}=15.8\pm1.3$~K for NGC\,1427A. Over the same aperture, our simultaneous fit of a two-component (warm+cold; see Sec.~\ref{app:ir}) dust model at the adopted distance in this work yields $\log(M_{\rm dust}/M_\odot)=6.60\pm0.25$, where the distance uncertainty contributes significantly to the error budget, at essentially the same temperature. The dust emission, however, extends significantly beyond the $D_{25}$ isophote considered by \citet{fullerHerschelFornaxCluster2014}. Our MIR-based estimate suggests an increase of up to a factor of $\sim2$ in the total dust mass when extending the aperture from the main body to the whole galaxy (from $\mu_0{+}1$ to $\mu_0{+}2$, still avoiding obvious contamination from bright background sources), implying a total dust mass of $\log(M_{\rm dust}/M_\odot)\simeq6.90\pm0.25$ when the full extent is considered. In what follows, we focus on the main body, where the dust geometry can be more directly compared to the stellar and ionized-gas distributions.

Adopting a thick exponential disk for the main body (half-light radius $R_e\!\simeq\!2.8$~kpc at $D\!\simeq\!17$~Mpc; \citealt{lee-waddellTidalOriginNGC1427A2018a}), an observed axial ratio $b/a\simeq0.65$, and an intrinsic thickness $q_0\simeq0.30$ appropriate for massive dwarfs \citep{sanchez-janssenThinDiscsThick2010, roychowdhuryIntrinsicShapesDwarf2013}, Holmberg's relation \citep{holmbergPhotographicPhotometryExtragalactic1958} implies an inclination of $i\simeq53^\circ$ for the stellar disk. Approximating for simplicity of this toy model that $R_e = D_{25}/2\simeq3$~kpc, the mean dust surface density is
\[
\Sigma_{\rm dust} \;=\; \frac{M_{\rm dust}}{\pi R_e^2}\;\simeq\;3.5\times10^{-5}\ {\rm g\,cm^{-2}}.
\]
Using a V-band mass absorption coefficient $\kappa_V\simeq8.6\times10^3\ {\rm cm^2\,g^{-1}}$ for Milky Way (MW) type grains \citep{draineInterstellarDustGrains2003}, the face-on and inclined optical depths are
\[
\tau_{V,\perp}=\kappa_V\,\Sigma_{\rm dust}\simeq0.30,\qquad
\tau_V=\frac{\tau_{V,\perp}}{\cos i}\simeq0.50.
\]
Two limiting geometries then bracket the attenuation \citep{disneyAreGalaxyDiscs1989, charlotSimpleModelAbsorption2000}: a foreground screen with $A_V^{\rm fs}=1.086\,\tau_V\simeq0.54$~mag, and a mixed slab with
\[
A_V^{\rm ms} \;=\; -2.5\,\log_{10}\!\left[\frac{1-e^{-\tau_V}}{\tau_V}\right]\;\simeq\;0.26\ {\rm mag}.
\]
Converting these to color excess using the Cardelli MW law ($R_V{=}3.1$; \citealt{cardelliRelationshipInfraredOptical1989}, C89 hereafter) and the Calzetti starburst law ($R_V{=}4.05$; \citetalias{calzettiDustContentOpacity2000}) to bracket the expected $R_V$ yields the expected $E(B{-}V)$ under idealized geometries, which is summarized in Table~\ref{tab:dust-expectations}.

\begin{table}[!h]
\centering
\caption{FIR-based expected $E(B{-}V)$ values in the main body under idealized geometries.}
\label{tab:dust-expectations}
\begin{tabular}{lcc}
\hline
Geometry & $E(B{-}V)$ \citepalias[][]{cardelliRelationshipInfraredOptical1989} & $E(B{-}V)$ \citepalias[][]{calzettiDustContentOpacity2000} \\
\hline
Foreground screen & $\simeq0.17$ & $\simeq0.13$ \\
Mixed slab        & $\simeq0.08$ & $\simeq0.06$ \\
\hline
\end{tabular}
\end{table}

Allowing for uncertainties in $\kappa_V$ ($\pm25\%$), in the intrinsic thickness $q_0$ (0.25-0.35), and in the dust extent, we estimate that the expected ranges are $E(B{-}V)_{\rm fs}\simeq0.05$-0.23 and $E(B{-}V)_{\rm ms}\simeq0.02$-0.11 for the main body.

The relative attenuation of stars and ionized gas is strongly geometry-dependent. In local starbursts and spiral disks, the nebular lines are typically more attenuated than the stellar continuum, with $E(B{-}V)_*\!\simeq\!0.44\,E(B{-}V)_{\rm gas}$ \citep[][\citetalias{emsellemPHANGSMUSESurveyProbing2022}]{calzettiDustOpacityStarforming2001}. In NGC\,1427A, we see the opposite. Inside the $\mu_0{+}1$ isophote, we measure median reddening of $\tilde E(B{-}V)_* \simeq 0.14$ and $\tilde E(B{-}V)_{\rm gas} \simeq 0.04$ (Sec.~\ref{sec:results-dust}; Fig.~\ref{fig:dust-properties}), i.e. $E(B{-}V)_*/E(B{-}V)_{\rm gas}\approx3.3$.\footnote{Mean reddening: $\langle E(B{-}V)_*\rangle \!\simeq\!0.17$; $\langle E(B{-}V)_{\rm gas}\rangle \!\simeq\!0.04$.} Placed against our FIR-based expectations (Table~\ref{tab:dust-expectations}), the stellar $E(B{-}V)$ is consistent with a foreground-screen-like configuration, whereas the nebular $E(B{-}V)$ lies at or below the mixed-slab prediction. The simplest configuration that reproduces both is one in which a substantial fraction of the dusty, cold ISM has been displaced in front of much of the older stellar population along our line of sight, while the currently ionized gas is mixed with the rest of the dust, with stellar feedback possibly clearing some of the dust, thereby further reducing the reddening \citep[see][for a numerical simulation of a similar geometry produced by RPS]{kronbergerInfluenceRampressureStripping2008}.

This extinction-based picture is fully consistent with the multi-phase kinematics. The \hi\ gas is systematically blueshifted relative to the stars (median $\sim\! -24$~km\,s$^{-1}$; Fig.~\ref{fig:deltaV_violinplot}) and forms extended tails (Sec.~\ref{subsec:results-morph-and-context}), indicating that gas and dust are being pushed toward the observer. The inferred dust-gas decoupling, with $E(B{-}V)_* \gg E(B{-}V)_{\rm gas}$, therefore provides independent evidence that ram-pressure stripping is acting with a strong line-of-sight component and has already started to affect the ISM within the main body, even though the velocity offset between stars and ionized gas remains modest. This geometry naturally explains the lack of a canonical jellyfish morphology in optical images: if most of the stripped gas and dust lies between us and the galaxy, we would not see bright \Halpha\ tentacles trailing away from the disk on the sky; instead, the stripped material acts as a foreground screen reddening the stellar continuum, while the nebular emission traces less-obscured pockets of ongoing star formation in the disk and inner tail.

\subsubsection{Gas-phase budget in the stripped system}
\label{sec:rps-gas-budget}

At face value, the multi-wavelength data show that NGC\,1427A currently hosts a substantial atomic reservoir but very little cold, CO-bright molecular gas, while its warm \Htwo\ and dust components remain prominent. Within the main body, MeerKAT yields an \hi\ mass of $\log_{10}(M_{\rm HI}/M_\odot)\simeq9.22$ (see Tab.~\ref{tab:globalprops}; \citealt{serraMeerKATFornaxSurvey2024}). In contrast, the ALMA Fornax survey detected no CO(1--0) emission from NGC\,1427A \citep{zabelALMAFornaxCluster2019a}. For non-detections in the ALMA Fornax survey, \citet{zabelALMAFornaxCluster2019a} estimate $3\sigma$ upper limits on $M_{\rm H_2}$ by assuming a Gaussian CO(1--0) line profile with ${\rm FWHM}=50~\mathrm{km\,s^{-1}}$ and adopting a metallicity- and sSFR-dependent CO-to-H$_2$ conversion factor $\alpha_{\rm CO}$ from \citet{accursoDerivingMultivariateACO2017} (their eq.~25), converted to $X_{\rm CO}$ following \citet{bolattoCOtoH2ConversionFactor2013}. For NGC\,1427A this yields an upper limit of $\log_{10}(M_{\rm H_2}/M_\odot)\!<\!7.42$ at the survey distance of 19.95~Mpc; rescaled to our adopted 17~Mpc distance, this becomes $\log_{10}(M_{\rm H_2}/M_\odot)\!<\!7.28$. For a galaxy with NGC\,1427A's stellar mass and star-formation rate, the \citet{accursoDerivingMultivariateACO2017} prescription implies an $\alpha_{\rm CO}$ of the same order as, but plausibly higher than, the canonical Galactic value ($\alpha_{\rm CO,MW}\simeq4.3~M_\odot\,\mathrm{\,[K\,km\,s^{-1}\,pc^{2}]^{-1}}$; \citealt{bolattoCOtoH2ConversionFactor2013}). Even allowing for such variations, the CO non-detection demonstrates that the cold, CO-bright molecular reservoir is small compared to the galaxy's substantial \hi\ mass. For metallicities comparable to those measured in the young star clusters by \citet{moraStarburstCoreGalaxy2015}, the inferred CO-based \Htwo\ upper limit would change by $\sim0.9$~dex.

A further caveat is spatial filtering. For the compact ALMA configurations used in the Fornax survey, the largest recoverable scale at 115~GHz is $\sim25\arcsec$, whereas the stellar and \hi\ disks of NGC\,1427A extend well beyond this and the bulk of the neutral gas (as well as \Halpha\ and FUV flux) is displaced toward the southwest relative to the optical center (Fig.~\ref{fig:rgb_meerkat_contours}). \citet{zabelALMAFornaxCluster2019a} explicitly note that, for this galaxy, diffuse CO emission on scales larger than $\sim\!25\arcsec$ may be resolved out by the interferometer. Thus, the quoted upper limit strictly applies to CO(1--0) structures on scales $\lesssim25\arcsec$ (i.e. $\sim\!2$--3~kpc at Fornax distance) within the primary beam; more extended, low-surface-brightness molecular gas associated with the outer, clumpy \hi\ distribution could in principle escape detection, especially in the southwestern regions where the neutral gas column peaks.

Even allowing for this possibility, combining the compact ALMA limit with the strong \hi\ reservoir still implies that any cold, CO-bright phase is subdominant: at face value the CO(1--0) non-detection gives $M_{\rm H_2,CO}/M_{\rm HI}\lesssim10^{-2}$, and even boosting $M_{\rm H_2,CO}$ by an order of magnitude to account for spatial filtering or enhanced metallicity would only bring $M_{\rm H_2,CO}/M_{\rm HI}$ into the $\sim0.1$ regime. For comparison, in typical late-type galaxies the molecular phase typically contributes $\sim15$\% of the total cold gas mass \citep{boselliEnvironmentalEffectsLateType2006}, and Virgo star-forming dwarfs show molecular-to-total gas fractions of $\sim\!9$-$38\%$ ($\simeq14$\% mean) despite the ongoing removal of their atomic and dust components by the cluster environment \citep{grossiStarformingDwarfGalaxies2016, boselliColdGasProperties2014}. NGC\,1427A thus lies toward the extreme, \hi-dominated end of the distribution of cluster star-forming dwarfs, with a cold CO-bright component that is at most comparable to, and likely lower than, the molecular fractions seen in Virgo and Herschel Reference Survey \citep{boselliHerschelReferenceSurvey2010} samples, even though its \hi\ disk remains massive and extended.

\subsubsection{Lessons from the presence of warm dust}
Independent constraints from mid-infrared spectroscopy suggest that a substantial warm molecular component is present. \citet{sivanandamTracingRampressureStripping2014} detect the \Htwo\ S(0) $J{=}2{\rightarrow}0$ line at 28\,$\mu$m in two IRS long-low apertures on NGC\,1427A, but find no corresponding S(1) $J{=}3{\rightarrow}1$ detection at 17\,$\mu$m. They derive a $3\sigma$ upper limit on the S(1)/S(0) flux ratio of $\lesssim0.4$, whereas the other galaxies in their sample have S(1)/S(0) in the range $\sim2$-6 and the SINGS galaxies show a median ratio of $3.1$ \citep{rousselWarmMolecularHydrogen2007}. In the SINGS sample, only three systems (NGC~24, NGC~1705, and NGC~4552) exhibit similarly low S(1)/S(0) ratios $<1.0$; all three have low ortho-to-para ratios, $0.5\lesssim{\rm OPR}\lesssim1.5$, and relatively cool warm-\Htwo\ components with $T\simeq80$-$130$~K \citep{rousselWarmMolecularHydrogen2007}. By analogy, the stringent S(1)/S(0) limit in NGC\,1427A most likely reflects a cool warm-\Htwo\ phase with an ortho-to-para ratio below the LTE value of 3. This is consistent with H$_2$ that formed in colder clouds and is now residing in a more diffuse, recently heated, low-density component where spin conversion is slow. For NGC\,1427A, \citet{sivanandamTracingRampressureStripping2014} infer a warm \Htwo\ mass lower limit of $\log_{10}(M_{\rm H_2,warm}/M_\odot)\gtrsim7.4$ (Table~\ref{tab:globalprops}), already exceeding the CO-based upper limit. Thus, even before accounting for any CO-dark component, the warm molecular phase contributes at least a few per cent of the \hi\ mass and plausibly dominates over the cold, CO-bright \Htwo.

The low S(1)/S(0) ratio and implied low OPR in NGC\,1427A therefore point to warm \Htwo\ that is either unusually cool or out of ortho-para equilibrium. In low-density gas, the ortho and para levels can remain decoupled for long times, so the OPR may stay ``frozen'' at its formation value rather than equilibrating to the high-temperature limit of 3 \citep{rousselWarmMolecularHydrogen2007, sivanandamTracingRampressureStripping2014}. This is qualitatively consistent with a scenario in which ram pressure and/or interaction-driven perturbations have already reached the ISM, dispersing or truncating the densest molecular clouds and leaving behind a diffuse, low-density warm-\Htwo\ component mixed with the extended \hi. In this picture, a substantial fraction of the molecular material may reside in extended, only moderately heated layers or RPS-rarefied filaments, while compact, well-shielded CO-bright clumps have a low filling factor.

\subsubsection{PAH signatures}

The mid-infrared aromatic bands provide independent clues on how the dust-bearing phase has been processed. Here we use ``PAH'' in the usual observational sense, while noting that these features may trace a broader population of very small aromatic carbonaceous grains rather than a single family of isolated PAH molecules. In NGC\,1427A the 6.2 and 11.3\,$\mu$m bands are clearly detected, but the 7.7\,$\mu$m feature is unusually weak in the spectrum: \citet{sivanandamTracingRampressureStripping2014} constrain the PAH(7.7)/PAH(11.3) flux ratio to $\lesssim\!1$, well below the typical values of $\sim\!3$--$4$ measured in normal star-forming galaxies \citep[e.g.,][]{smithMidInfraredSpectrumStarforming2007, rousselWarmMolecularHydrogen2007}. Models and observations of PDRs show that mid-IR PAH ratios depend on charge state, size distribution, molecular structure, and radiation-field conditions; in broad terms, the 6.2 and 7.7\,$\mu$m bands are enhanced in more ionized PAH populations, whereas the 11.3\,$\mu$m band is stronger for more neutral aromatic grains \citep[e.g.][]{draineInfraredEmissionInterstellar2001, flageySpitzerIRACISOCAM2006, ogleJetpoweredMolecularHydrogen2010, maragkoudakisProbingSizeCharge2020}. 

NGC\,1427A also shows a high ${\rm PAH}(6.2)/{\rm PAH}(7.7)\gtrsim0.8$, implying that the 7.7\,$\mu$m complex is weak relative to both the 6.2 and 11.3\,$\mu$m bands. This points to an unusual aromatic spectrum, but it does not uniquely determine the underlying grain-size distribution. Fine-structure line diagnostics show that the radiation field is not unusually hard: the [Ne\,\textsc{iii}]~15.6\,$\mu$m to [Ne\,\textsc{ii}]~12.8\,$\mu$m ratios are $\mathrm{[Ne\,III]/[Ne\,II]}\lesssim0.1$ for the global extraction and $\lesssim0.3$ in the brightest region \citep[][their Table~3]{sivanandamTracingRampressureStripping2014}, values characteristic of normal star-forming galaxies and well below those of AGN hosts with similarly depressed 7.7/11.3\,$\mu$m ratios \citep[e.g.][]{smithMidInfraredSpectrumStarforming2007}. 
A plausible interpretation is therefore that the weak 7.7\,$\mu$m complex reflects a processed aromatic/small-grain population rather than an unusually hard radiation field. This processing could involve changes in the PAH charge balance, changes in the grain-size distribution, or altered excitation and survival of small grains in stripped or shocked gas \citep[e.g.][]{egorovPHANGSJWSTFirstResults2023, chastenetPHANGSJWSTFirstResults2023}.
However, without access to the 3.3\,$\mu$m feature, which provides a stronger constraint on the smallest aromatic grains (as clearly shown in Fig.~7 of \citealt{maragkoudakisProbingSizeCharge2020}), the low ${\rm PAH}(7.7)/{\rm PAH}(11.3)$ ratio should not be interpreted as sufficient evidence for selective destruction of the smallest PAHs \citep[see also][]{maragkoudakisPolycyclicAromaticHydrocarbon2023, tarantinoJWSTCapturesGrowth2025}.

The spatial information from the IRAC 8\,$\mu$m map supports this picture. In the bright main body, the 8\,$\mu$m emission wraps around the \Halpha\ knots, as expected for PAH-bearing material in and around dusty \ion{H}{ii} regions and adjacent PDRs. By contrast, some of the highest surface-brightness \Halpha\ clouds on the southeastern edge show little 8\,$\mu$m emission \citep[][and Fig.~\ref{fig:MIR-FIR-dust}]{sivanandamTracingRampressureStripping2014}, coincident with regions of elevated \Halpha\ velocity dispersion (Fig.~\ref{fig:ngc1427a_15panel}). The outermost star-forming regions, therefore, appear relatively faint in aromatic/small-grain emission compared to \Halpha. This is broadly consistent with local processing of the dust-bearing ISM by stellar feedback, shocks, and/or stripping. Future narrow- and medium-band JWST NIR and MIR observations could test this interpretation by mapping the 3.3, 7.7, and 11.3\,$\mu$m features at sub-kpc scales and comparing them with the MUSE and UVIT tracers of star formation and ionization \citep[cf.][]{tarantinoJWSTCapturesGrowth2025}.

Taken together, the large \hi\ reservoir, stringent CO(1--0) upper limit, and bright but cool warm-\Htwo\ emission indicate that the ISM in and around NGC\,1427A is dominated by atomic gas and warm molecular material, while any cold CO-bright molecular component is comparatively limited. This does not require the absence of cold \Htwo: in low-metallicity or porous media, \Htwo\ can survive in regions where CO is faint or photodissociated, producing a substantial CO-dark molecular component \citep[e.g.][]{bolattoCOtoH2ConversionFactor2013, maddenTracingTotalMolecular2020}. The measured cold dust mass can be reconciled with the CO(1--0) non-detection. A global dust reservoir does not guarantee that the dust is distributed in compact, high-$A_V$ structures with sufficient local shielding to maintain an extended CO-bright phase, especially if the elevated star formation over the past $\sim$Gyr has increased the porosity of the ISM and reduced the filling factor of well-shielded molecular clouds.

The low S(1)/S(0) ratio and likely non-equilibrium OPR point to warm \Htwo\ residing in a diffuse, low-density component that still preserves the spin distribution of its colder progenitor clouds. This phase balance is consistent with environmental processing acting on a low-mass system: ram pressure, repeated stellar feedback over the recent SFH, and/or a recent tidal perturbation can disperse dense gas, heat or shock molecular material, and reduce the filling factor of well-shielded CO-bright clouds. In this configuration, the ISM would be more susceptible to further stripping and less efficient at sustaining ongoing star formation, consistent with the declining SFRs presented in Sect.~\ref{sec:sfr_results}.


\subsection{Tidal perturbations and the FCC\,229 fly-by scenario}\label{sec:flyby_localised}
Beyond the global signatures of ram-pressure stripping discussed above, NGC\,1427A exhibits a set of more localized, non-collisional features that are difficult to attribute to ICM-ISM interactions alone and instead point to a recent localized tidal perturbation.

\subsubsection{Misalignment and global morphology}
The first indication is a kinematic tilt between the stellar rotation axis and the cold and warm gas: both the \hi\ and ionized-gas velocity fields are systematically misaligned with respect to the stellar component, with the gas disk clearly reoriented relative to the underlying stars. Such a configuration is naturally produced when a gravitational encounter torques the gaseous disk while leaving the more centrally concentrated stellar body largely intact, as seen in simulations of dwarf-dwarf interactions \citep[e.g.][]{starkenburgDarkInfluencesII2016, zengKinematicMorphologyLowmass2024}. In addition, the highest surface-density \hi, the bulk of the ionized gas, and the warm-dust emission are all displaced toward the southwest with respect to the stellar mass distribution. 
This off-center gas and dust concentration is difficult to obtain purely through interactions with the ICM wind, which tends to act more immediately on the outer disk \citep[e.g.][]{kronbergerInfluenceRampressureStripping2008}, but follows naturally from an off-axis tidal perturbation that displaces the gaseous disk and compresses it locally as the perturber passes. We discuss below the potential effects at play in NGC\,1427A.

\subsubsection{Cluster tides, pre-processing, and galaxy--galaxy perturbations}

The cluster-tide scenario of \citet{mastropietroTaleTwoTails2021} is an important alternative to consider. In that framework, the NE--SW stellar and gaseous extension can arise from the combined action of the Fornax tidal field, galaxy rotation, orbital curvature, and RPS near pericenter, without requiring a specific nearby perturber. We distinguish this scenario from a broader form of environmental pre-processing: NGC\,1427A may have been affected by interactions during infall through the Fornax environment without the central cluster potential being the dominant tidal agent. This distinction is important because mergers and close galaxy--galaxy encounters are expected to occur preferentially along filaments and cluster outskirts, where relative velocities are lower than in virialized cluster cores \citep[e.g.][]{dulcienCaughtWebGalaxy2026}.

The deeper MeerKAT analysis of \citet{serraMeerKATFornaxSurvey2024} shifts the balance toward this latter picture. Compared to the older ATCA-based view, MeerKAT reveals a longer, more structured, and kinematically complex \hi\ tail. \citet{serraMeerKATFornaxSurvey2024} argue that the global Fornax tidal field is unlikely to dominate the observed disturbance, and instead favor a high-speed encounter or advanced merger that tidally disturbed the gas before RPS shaped the extended tail. That event would have occurred on a timescale of at least a few $\times 10^8$~yr, and therefore should not be identified one-to-one with the more recent interaction discussed below.

Our MUSE and UVIT results add a shorter-timescale constraint. The FUV/\Halpha\ morphology, the off-center young component, and the present gas--star kinematic decoupling point to an additional recent, localized perturbation over the past $\sim30$--$100$~Myr. We therefore favor a two-stage interpretation: an older galaxy--galaxy perturbation or merger, together with RPS, pre-conditioned the outer gas and produced the large-scale \hi\ disturbance on timescales of at least a few $10^8$~yr, while a more recent mild fly-by, with FCC\,229 as the favored projected candidate, may have modulated the current star-forming ISM over the past $\sim30$--$100$~Myr.

\subsubsection{FUV vs. \Halpha: Location and timing of star formation episodes}

A second line of evidence comes from the relative distribution of FUV and \Halpha\ emission. The only region that is clearly FUV-bright yet \Halpha-faint is the diffuse northeast extension (Fig.~\ref{fig:FUV-HA-8UM}), located immediately east of the northern clump. This morphology is consistent with a fading star-formation episode on $\sim30$--$100$~Myr timescales, older than the currently FUV- and \Halpha-bright sites concentrated in the main body. The absence of comparable FUV-only features elsewhere argues against a global trigger and instead points to a localized perturbation acting preferentially on the northeastern outer disk.

A schematic fly-by scenario can account for this configuration: a close passage on the near, northeastern side of the disk could have compressed the outer gas layers and triggered star formation there. In this picture, the northern clump could mark the primary impact/compression site where star formation remains ongoing (traced by \Halpha), while the adjacent diffuse FUV-only emission would reflect an older, already fading episode. As the disk rotated and the perturber continued along its orbit toward the current projected southwest direction, the original impact region would not have had time to move far in azimuth. {Adopting as reference the stellar rotation speed of $\sim\!20$ km s$^{-1}$ at a radial distance of 3\,kpc (see Fig.~\ref{fig:ngc1427a_15panel}), the northern FUV-bright region in Figure~\ref{fig:ngc1427a-ngfs-isophotes} would have azimuthally moved in $30\!-\!100$ Myr counterclockwise by $12^\circ\!-\!40^\circ$, i.e. less than a full quadrant.}

The stellar population fits in the outer disk include a young, comparatively metal-poor component, suggesting that part of the fuel could have been low-metallicity gas, either pre-existing at large radii or redistributed during an interaction. In the schematic recent fly-by picture ($\lesssim$100~Myr; Fig.~\ref{fig:scenario-cartoon}), the most plausible perturbers are the nearby dwarfs discussed earlier (Table~\ref{tab:nearby-dwarfs}), with FCC\,229 being the most compelling candidate in projection for this recent localized component. {On longer timescales of a few $10^8$~yr, additional galaxies at larger projected separations (e.g. FCC\,248, FDS6\,100b, and FCC\,247) provide plausible earlier interaction candidates that could have contributed to the prior gas decoupling required to produce the extended H\,I tail and detached clouds \citep[][]{serraMeerKATFornaxSurvey2024, Kleiner2025}; however, their dynamical role remains uncertain in the absence of 3D orbit constraints.} Ram-pressure stripping can also enhance star formation through compression on the wind-facing side, but such enhancements typically act on a global scale \citep[e.g.][]{kronbergerInfluenceRampressureStripping2008, steinhauserSimulationsRampressureStripping2016}. We therefore interpret the northeastern FUV/\Halpha\ mismatch as the imprint of a locally triggered episode whose ionizing population has largely faded, superposed on the more global disturbance over both the CGM and ISM.

We stress that this fly-by scenario is intended as a schematic, non-collisional complement to the ram-pressure picture; quantitative tests will require tailored hydrodynamical modeling and a more complete reconstruction of the interaction geometry and disk rotation curve.

\subsubsection{Candidate perturber FCC 229}
\label{sec:perturber-fcc229}

Among the nearby dwarfs discussed in Table~\ref{tab:nearby-dwarfs}, FCC\,229 is the most plausible candidate for a recent tidal perturber in projection, lying to the southwest of NGC\,1427A (Fig.~\ref{fig:ngc1427a-ngfs-isophotes}). In deep NGFS imaging, its stellar body appears largely regular, with no obvious shells, streams, or strong asymmetries (Fig.~\ref{fig:FCC229_morph}). This disfavors a very strongly disruptive encounter, but the absence of clear stellar debris is not a stringent constraint for dwarf spheroidals, which can remain photometrically smooth under tidal forcing for many dynamical times, particularly if dark-matter dominated (e.g. \citealt{fellhauerModellingDynamicalEvolution2008}). We therefore treat the current morphology as consistent with a tidally mild fly-by, and revisit this interpretation below in light of the nuclear structure of FCC\,229.

The optical luminosity contrast should not be interpreted directly as the total encounter mass ratio. FCC\,229 has $M_\star\simeq10^{7.6}\,M_\odot$, compared to $M_\star\simeq10^{9.3}\,M_\odot$ for NGC\,1427A, corresponding to a stellar-mass ratio of order $50{:}1$. However, for dwarf galaxies, the stellar-to-halo mass relation is steep and uncertain, and the total pre-infall halo mass contrast can be substantially less extreme than the stellar contrast alone suggests \citep[down to order $5:1$; e.g.][]{mosterGalacticStarFormation2013, readStellarMasshaloMass2017}. Even if FCC\,229 has been tidally stripped, a dark-matter-dominated remnant could still perturb the outer gas disk of NGC\,1427A without producing a strongly disrupted stellar morphology in FCC\,229 itself. Simulations of dwarf tidal stripping show that the inner stellar body of dark-matter-dominated dwarfs can remain comparatively regular after substantial mass loss \citep[e.g.][]{penarrubiaTidalEvolutionLocal2008, smithImpactGalaxyHarassment2013, erraniStructureKinematicsTidally2022}. Thus, FCC\,229 is not ruled out by the optical mass contrast alone, although the interaction must have been mild and/or primarily affected the extended gaseous component.

In this configuration, enhanced recent star formation in NGC\,1427A could arise primarily through tidal compression of its outer gas layers, inducing non-axisymmetric structure and promoting cloud collapse even in the absence of direct gas--gas collisions. A secondary possibility is that FCC\,229 carried some gas at pericenter, allowing additional hydrodynamic interaction with the outer \hi\ disk of NGC\,1427A and potentially contributing low-metallicity fuel. However, the present-day optical morphology of FCC\,229 and the lack of an obvious associated starburst argue that, if FCC\,229 was gas-bearing, it likely contained only a modest and/or diffuse reservoir at the time of the encounter, which is not constrained by existing data.  Conversely, if the fly-by involved substantial gas--gas interaction at relative velocities of a few $10^2$~km~s$^{-1}$, one might expect stronger, more conspicuous star formation signatures and/or kinematic components tied to the perturber; testing this requires tailored simulations and deeper gas observations.

A simple timescale estimate supports the plausibility of a recent encounter. Adopting the projected separation between FCC\,229 and NGC\,1427A, $d\simeq20$~kpc, and a characteristic relative velocity of order the Fornax velocity dispersion, $\Delta v \sim \sigma \simeq 350$~km~s$^{-1}$ \citep{Schuberth10, chaturvediFornaxClusterVLT2022}, yields $t_{\rm fb}\sim d/\Delta v \simeq 55$~Myr. Given the large uncertainty in the true 3D separation and relative velocity, this should be regarded as an order-of-magnitude consistency check, but it matches the $\sim30$--$100$~Myr timescale implied by the fading FUV-only episode (Sect.~\ref{sec:disc-sfr-sfh}; Fig.~\ref{fig:stellar-pdfs-apertures}).

The internal structure of FCC\,229 is compatible with, but not diagnostic of, mild tidal perturbation. Morphological modeling reveals a compact nuclear component ($m_{i'}\simeq23.5$) offset by $\sim1\arcsec$ ($\sim80$--$100$~pc) from the center of the best-fitting S\'ersic profile in $g'$ and $i'$ (Fig.~\ref{fig:FCC229_morph}). In low-mass galaxies, nuclear star clusters can remain displaced for many dynamical times after tidal perturbations because of long dynamical-friction timescales in low-density (and potentially cored) potentials \citep[e.g.][]{bellovaryOriginsOffcentreMassive2021, Fahrion21, poulainEvidenceStarCluster2025}. The putative NSC has colors consistent with the inner galaxy light, arguing against an extremely young nuclear starburst. While not definitive on its own, the combination of a modest nuclear offset with an otherwise regular stellar morphology is compatible with FCC\,229 having experienced a recent, tidally mild encounter.

Ultimately, the gas content and dynamical impact of FCC\,229 remain weakly constrained with current data. Deeper H\,I and CO observations and spectroscopy of FCC\,229 are required to test whether it carried gas at pericenter and to place tighter constraints on the interaction strength and timing.

\subsection{Star formation rate properties of NGC 1427A}
\label{sec:disc-sfr-sfh}
{The star formation history of NGC\,1427A is best interpreted by combining tracers that probe different timescales. \Halpha\ emission is sensitive to the massive O-type stars and therefore traces star formation over the last $\sim\!10$~Myr (with mean age $\sim\!3$~Myr). The FUV continuum responds to stars of somewhat lower mass and integrates star formation over the past $\sim\!100$~Myr (with mean age $\sim\!10$~Myr; \citealt{kennicuttStarFormationMilky2012}). Our pPXF stellar-population modeling infers older stellar population ages ($\log[{\rm Age}]\lesssim10.4$, see Fig.~\ref{fig:stellar-pdfs-apertures}) and includes youngest templates with ages of $\sim30$~Myr.}

\begin{figure}
    \centering
    \includegraphics[width=\linewidth]{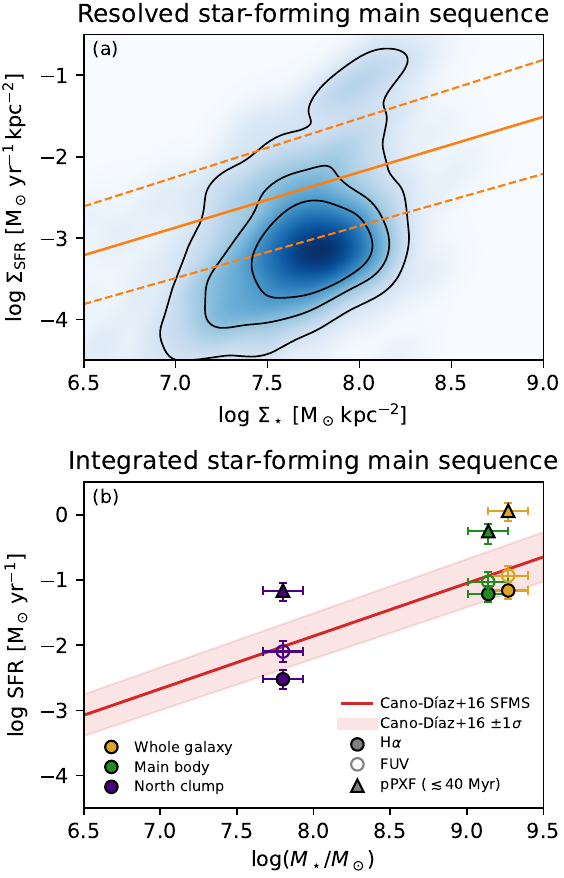}
    \caption{(a) Resolved SFMS (KDE): dust-corrected \Halpha-based $\Sigma_{\rm SFR}$ versus $\Sigma_\star$ for Voronoi bins. Solid and dashed lines show the best-fit resolved SFMS and $\pm1\sigma$ envelope from \citet{cano-diazSPATIALLYRESOLVEDSFR2016}. (b) Integrated SFMS for the whole galaxy, main body, and northern clump compared to the local SFMS \citep{cano-diazSPATIALLYRESOLVEDSFR2016}. Filled circles: \Halpha-based SFRs (App.~\ref{app:DAPmethods}); open circles: FUV-based SFRs (App.~\ref{app:uvit}). Triangles show the pPXF-inferred mean SFR in the youngest SFH bin (30~Myr template, which integrates over the past $\sim$40~Myr).
}
    \label{fig:sfms-combined}
\end{figure}

{The resolved star-formation main sequence (SFMS) provides a stringent, sub-kpc diagnostic of how efficiently different patches of the NGC\,1427A disk are converting gas into stars at fixed underlying stellar surface density, and therefore whether environmental processing is acting as a smooth “global dimmer” or as a highly structured, local phenomenon. The kernel-density estimate distribution of the \Halpha-based SFR surface density relation $\Sigma_{\rm SFR}$-$\Sigma_*$ in Figure~\ref{fig:sfms-combined}a peaks at $\gtrsim\!1\sigma$ below the SFMS of \cite{cano-diazSPATIALLYRESOLVEDSFR2016}, but is noticeably skewed toward enhanced $\Sigma_{\rm SFR}$. This implies that widespread regions are forming stars inefficiently, while a smaller subset still hosts locally boosted, high-$\Sigma_{\rm SFR}$ clumps. In the fly-by + first-infall picture, this combination is expected as a recent satellite encounter can torque/displace the gaseous disk and produce localized compression (feeding the high-$\Sigma_{\rm SFR}$ tail), while the onset of ICM interaction during first infall simultaneously strips/shreds the ISM and suppresses star formation over much of the disk (shifting the mode of the distribution below the ridge line).}

\subsubsection{SFRs from multi-timescale indicators}
Viewed together, the three SFR tracers (\Halpha, FUV, pPXF stellar population modeling) reveal a coherent temporal sequence of the star formation history (SFH). The youngest pPXF bin indicates that the most recent burst was $\sim\!3\sigma$ above the SFMS at the recently formed $M_*$ (see Fig.~\ref{fig:sfms-combined}), i.e., NGC\,1427A experienced a strong star-formation burst when integrated over the last $\sim\!30\!-\!40$~Myr. The FUV-based SFR places the galaxy on or slightly below the local SFMS, while the \Halpha-based estimate lies $\sim\!1\sigma$ below it. This ordering, therefore, points to a system whose SFR has peaked within the last few $10$~Myr and is declining on a similar timescale. In combination with the evidence for a displaced, disturbed ISM, this trend suggests that NGC\,1427A is entering an environmentally driven quenching phase.

A caveat in comparing SFR$_{\rm FUV}$ and SFR$_{\rm H\alpha}$ is the dust geometry. In this work we assume $E(B{-}V)_{\rm FUV}\!=\!E(B{-}V)_{\rm gas}$ (Sec.~\ref{sec:sfr_results}; App.~\ref{app:uvit}), i.e., the young stellar continuum and nebular gas suffer the same dust attenuation. {If ram-pressure stripping and stellar feedback clear dust and gas more efficiently along sightlines to the very youngest, \Halpha-emitting regions than toward the slightly older FUV-bright populations, then the Balmer decrement would trace a lower effective attenuation than the FUV continuum. In that case, our FUV attenuation would be overestimated and ${\rm SFR}_{\rm FUV}$ biased low, implying that the true decline from pPXF $\rightarrow$ FUV $\rightarrow$ \Halpha\ is at least as strong as observed, and possibly more pronounced.}

\subsubsection{Global and resolved star-forming main sequence}
On integrated scales, the three apertures (whole galaxy, main body, and northern clump) follow a consistent pattern relative to the global SFMS (Fig.~\ref{fig:sfms-combined}b). As summarized in Table~\ref{tab:globalprops} and Sec.~\ref{sec:sfr_results}, the FUV-based SFRs place NGC\,1427A on or only slightly below the local SFMS of \citet{cano-diazSPATIALLYRESOLVEDSFR2016}, whereas the \Halpha-based values sit $\sim\!1\sigma$ below the relation. The main body follows the same behavior, and the northern clump shows a slightly stronger suppression in \Halpha, consistent with enhanced stripping in the outskirts. Visually, the fact that FUV-based SFRs (open symbols in Fig.~\ref{fig:sfms-combined}b) lie systematically closer to the SFMS ridge line than \Halpha-based SFRs (filled symbols) provides a compact representation of the temporal SFR decline mentioned above. The pPXF-based SFRs (triangles) sit substantially above the SFMS, reflecting the much higher SFR when integrated over the last $\sim$40 Myr compared to the present-day \Halpha\ and FUV indicators.

{The fact that the resolved offset is larger than the integrated one further indicates that the environmental impact is patchy (and therefore partially "averaged out" in global apertures), consistent with the integrated behavior where FUV places the galaxy near the SFMS while \Halpha\ lies $\sim\!1\sigma$ lower, which implies a recent decline on $\lesssim\!100$~Myr timescales superposed on spatially intermittent and interaction-triggered star formation. Part of this offset may simply reflect that the reference resolved SFMS from \cite{cano-diazSPATIALLYRESOLVEDSFR2016} is dominated by more massive, more metal-rich disks, which can exhibit higher star-formation efficiencies at fixed $\Sigma_*$. Nonetheless, this systematic does not affect the central result, i.e., the distribution's shape in Figure~\ref{fig:sfms-combined}a remains a robust signature of spatially heterogeneous star formation in NGC\,1427A.}

\subsubsection{Effects on star formation from cluster infall}
On longer timescales, the pPXF-based SFH indicates an underlying old stellar population plus a prolonged episode of moderately elevated star formation over the last $\sim\!1$~Gyr at $\dot{M}_*\!\approx\!(3$-$4)\times10^{-2}~M_\odot\,\mathrm{yr^{-1}}$. The youngest pPXF bin and the FUV/\Halpha\ tracers together reveal a strong recent burst in the last $\sim30$-40~Myr followed by a current decline, i.e., a transition from burst to quenching. This kind of gigayear-scale enhancement with shorter-lived, spatially migrating starbursts is reminiscent of the behavior seen in nearby dwarf galaxies \citep[e.g.][]{mcquinnNATURESTARBURSTSII2010a}, and is echoed in NGC\,1427A by the mix of FUV-bright/\Halpha-faint and FUV-faint/\Halpha-bright regions across the galaxy (Fig.~\ref{fig:FUV-HA-8UM}).

These internal SFH signatures are naturally embedded in a larger-scale infall scenario. The accretion of galaxies into clusters is expected to be anisotropic, occurring preferentially along filaments and planar structures within the surrounding dark matter distribution. Recent work on the spatial and kinematic distribution of Fornax globular clusters in galaxies and in the intra-cluster population reveals coherent structures, some filament-like, extending toward the southeastern region from which NGC\,1427A may be approaching \citep[e.g.][]{dabruscoSpatialDistributionIntraCluster2025,chaturvediFornaxClusterVLT2022}. NGC\,1427A's elevated SFR over the past Gyr could reflect the period during which it was entering and traveling along a filament, with an increased occurrence of encounters with other dwarfs, boosting star formation \citep{stierwaltTiNyTITANSROLE2015a} or experiencing pre-processing in general. Internal secular mechanisms (e.g., bar-driven inflows) could in principle contribute, but bars are expected to be uncommon at NGC\,1427A’s low stellar mass, and we do not see an obvious bar signature in the optical morphology \citep[e.g.][]{mendez-abreuLackStellarBars2011}.

In this view, the $\sim$Gyr-scale elevated SFR marks an earlier infall/pre-processing stage that plausibly set the initial conditions for the more recent, shorter-timescale evolution discussed in Sects.~\ref{sec:rps-gas-budget} and~\ref{sec:flyby_localised} (Fig.~\ref{fig:scenario-cartoon}).

\subsection{Stellar metallicity in context}

\begin{figure}
    \centering
    \includegraphics[width=0.9\linewidth]{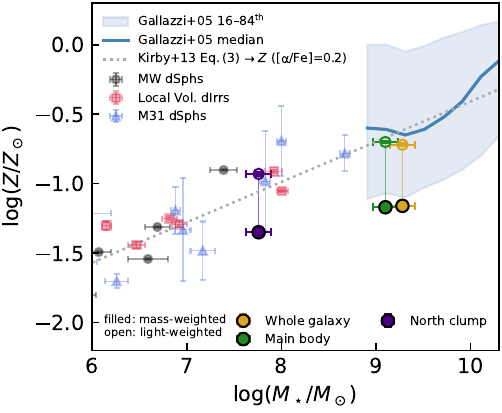}
    \caption{\textbf{Stellar mass-stellar metallicity plane.}
    Blue line and band: SDSS stellar mass-metallicity relation from \citet{gallazziAgesMetallicitiesGalaxies2005}.
    Gray dotted line: dwarf-galaxy relation from \citet{kirbyUniversalStellarMassStellar2013a}, converted to total metallicity and stellar mass as discussed in the text.
    Colored symbols: NGC\,1427A measurements in three apertures (gold = whole galaxy; green = main body; purple = northern clump).
    Filled circles show pPXF mass-weighted metallicities $\langle\mathrm{[Z/H]}\rangle_{\rm MW}$; open circles show light-weighted values $\langle\mathrm{[Z/H]}\rangle_{\rm LW}$, with vertical connectors.}
    \label{fig:mass_met_relation}
\end{figure}

Figure~\ref{fig:mass_met_relation} places the stellar metallicities of NGC\,1427A on the stellar mass-metallicity plane relative to two widely used, but methodologically distinct, reference relations: i) the light-weighted SDSS relation of \citet{gallazziAgesMetallicitiesGalaxies2005} and ii) the Local Group dwarf relation of \citet{kirbyUniversalStellarMassStellar2013a}, converted from luminosity to stellar mass and from [Fe/H] to total metallicity using a constant $[\alpha/\mathrm{Fe}]\!=\!0.2$~dex, following \citet{romero-gomezSAMIFornaxDwarfsSurvey2023}. For each of our apertures (whole galaxy, main body, northern clump), we plot both light-weighted (LW) and mass-weighted (MW) values derived with pPXF.

Across all apertures we find $\langle\mathrm{[Z/H]}\rangle_{\rm LW}\!>\!\langle\mathrm{[Z/H]}\rangle_{\rm MW}$ by $\sim\!0.4\!-\!0.5$~dex (Table~\ref{tab:globalprops}). This offset reflects a luminous young component superposed on an old, more metal-poor stellar body, consistent with the age structure discussed in Sec.~\ref{sec:stellar_pop_properties}. For the whole-galaxy aperture, NGC\,1427A's light-weighted metallicity falls slightly below the median SDSS mass-metallicity relation of \citet{gallazziAgesMetallicitiesGalaxies2005} and the \citet{kirbyUniversalStellarMassStellar2013a} Local Group dwarf sequence, which are mutually consistent at this mass. The outer regions of the galaxy, including the northern clump, are on average more metal-poor.

The northern clump is indeed slightly more metal-poor by $\sim0.2$~dex (in both LW and MW values), but still lies within the scatter of dwarf mass-metallicity relations. Together with its regular stellar and gas kinematics and the lack of enhanced velocity dispersion, this disfavors the interpretation of the clump as a separate dwarf currently merging with NGC\,1427A. Instead, its metallicity and kinematics are compatible with it being an off-center star-forming region within the same system, plausibly linked to the recent fly-by discussed in Section~\ref{sec:flyby_localised}.

Independent constraints from star clusters support this picture of enrichment. \citet{georgievOldGlobularCluster2006} find old globular clusters around NGC\,1427A with typical sub-solar metallicities, indicating that the early assembly of the system proceeded from metal-poor gas, as in other low-mass dwarfs. In contrast, \citet{moraStarburstCoreGalaxy2015} reported near-solar metallicities for young star clusters associated with the current starburst, with a complete picture of the star clusters presented by \citet{Fahrion2026a}. Our pPXF inferred SFHs show the two components (see marginal metallicity histograms in Fig.~\ref{fig:stellar-pdfs-apertures}), as well as the marked differences between MW and LW measurements. 
This combination suggests that NGC\,1427A remained overall metal-poor for its mass, likely due to a history of metal-loaded outflows in the field phase and, more recently, formed stars in star cluster complexes that efficiently retained their metals, and dominated the fraction of star formation as suggested by \citet{moraStarburstCoreGalaxy2015}.

Methodological differences between the reference relations and our pPXF-based estimates, as well as uncertainties in the [Fe/H]-to-[Z/H] conversion, may shift the absolute metallicity scale by $\sim0.1$~dex, but do not alter this qualitative scenario.

\section{Summary and Conclusions}
\label{sec:conclusions}
We presented a spatially resolved, multi-phase view of the Fornax dwarf NGC\,1427A, combining a custom-reduced four-field VLT/MUSE mosaic with optical, FUV, IR, and MeerKAT \hi\ data. Our aim was to characterize how ram-pressure stripping and tidal perturbations jointly transform a gas-rich, star-forming dwarf during first infall into the Fornax cluster. In the following, we summarize our work and provide conclusions.

\begin{enumerate}
  \item \textbf{NGC\,1427A is a resolved benchmark for early dwarf transformation in Fornax.}
  NGC\,1427A is a gas-rich, actively star-forming dwarf galaxy with a disturbed ``cometary'' stellar body, a long one-sided \hi\ tail, a secondary short \hi\ feature, and an extended low-surface-brightness envelope \citep[e.g.][]{munozUNVEILINGRICHSYSTEM2015a, iodiceFornaxDeepSurvey2016}. Together with the growing MeerKAT census of disturbed, gas-rich Fornax dwarfs \citep[][]{serraMeerKATFornaxSurvey2023a, kleinerMeerKATFornaxSurvey2023a}, it anchors a nearby sample of systems caught during transformation. Its surface brightness and our spatially complete MUSE mosaic make it an unusually clean resolved benchmark for studying the onset of cluster-driven transformation in low-mass galaxies. Our four-field MUSE mosaic provides contiguous coverage of the main body and outskirts with homogeneous reduction and robust sky subtraction. Combined with AstroSat/UVIT FUV, NGFS optical imaging, Spitzer/Herschel dust and gas tracers, and MeerKAT \hi\ observations, we map stellar populations, ionized gas, neutral gas, and dust, deriving stellar and gas kinematics, extinction, ages, metallicities, and $\Sigma_{\rm SFR}$ in a self-consistent way.
  
  \item \textbf{Global environmental picture: ram-pressure stripping plus tides.}
  NGC\,1427A shows a clear misalignment between the stellar and gaseous kinematics. A simple characterization of the large-scale velocity gradients (see Appendix~\ref{app:kinpa}) yields kinematic major-axis position angles of $63.6^\circ \pm 1.4^\circ$ for the stars, $91.4^\circ \pm 0.5^\circ$ for H$\alpha$, and $102.0^\circ \pm 1.8^\circ$ for \hi. These values confirm that both gas phases are significantly tilted with respect to the stars, while remaining more nearly aligned with each other.
  Furthermore, the \hi\ gas is globally blueshifted and the \Halpha\ velocities are skewed to the blue, with the median marginally offset from the stars, indicating that a substantial fraction of the disk gas is currently being removed with a significant line-of-sight component toward us. Comparing stellar and nebular attenuation with IR dust tracers shows that dust and gas are displaced in front of the old stellar body, but remain tightly coupled to the ionized gas where the surface brightness is high. This is expected if the collisional components in the inner disk and ISM are now being compressed and stripped along the line of sight.

  The long, mostly starless MeerKAT \hi\ tail lies roughly in the plane of the sky and points toward the anti-cluster-center direction. Together with the modest cluster tidal field, this identifies RPS by the Fornax ICM as the dominant driver of the extended gas removal from the circumgalactic and outer-disk regions over the past few $\times100$~Myr. As argued by \citet{serraMeerKATFornaxSurvey2024}, the absence of a strongly compressed leading edge in projection during this earlier, more plane-of-sky phase suggests that an additional mechanism had already decoupled and stirred the outer gas, making it easier for the ICM to peel off. Our MUSE-based dust and kinematic diagnostics add a complementary view of the current orbital phase: while the outer \hi\ tail traces gas removed earlier along a stripping direction with a stronger plane-of-sky component, the displaced dusty and ionized ISM in the inner body indicates that the present stripping geometry has acquired a substantial line-of-sight component toward the observer. In this sense, the outer tail and the dusty inner ISM do not trace the same instantaneous geometry, but different stages of a curved infall through the Fornax potential. In this work, we identify and characterize a plausible tidal perturber, summarized in the next item.

  \item \textbf{A localized tidal perturbation: FCC\,229 as the favored recent perturber.}
  Beyond RPS alone, there are indications for a global gas-star kinematic misalignment, the southwestern pile-up of high-density \hi, ionized gas, and warm dust offset from the stellar mass peak, and the exclusively northeastern FUV-bright/H$\alpha$-faint extension. Together with the arguments of \citet{serraMeerKATFornaxSurvey2024} for early circumgalactic disturbance, these observations point to an additional tidal perturbation. We have considered the cluster-tide scenario of \citet{mastropietroTaleTwoTails2021}; however, the deeper MeerKAT analysis favors a galaxy--galaxy tidal perturbation or merger plus RPS over a dominant role for the global Fornax tidal field. Within this broader picture, FCC\,229 is the favored nearby candidate for the recent localized component, based on its projected alignment, crossing-time estimate, and internal structure.

  \item \textbf{Star formation: a recent burst followed by an early decline, consistent with the onset of quenching.}
  Multi-timescale SFR tracers (pPXF, FUV, and H$\alpha$) reveal a coherent temporal sequence. The youngest star-formation history bin ($\sim\!30\!-\!40$~Myr from pPXF) implies a recent SFR peak $\sim\!3\sigma$ above the star-forming main sequence (SFMS) at the current stellar mass. The FUV-based SFR lies on or slightly below the local SFMS, with the caveat that our assumption $E(B{-}V)_{\rm FUV}\!=\!E(B{-}V)_{\rm gas}$ may underestimate the true FUV attenuation if RPS clears dust preferentially around the youngest \Halpha-emitting regions. The H$\alpha$-based SFR falls $\sim\!1\sigma$ below the SFMS, indicating that the instantaneous SFR has already declined from the recent peak. 
  
  Resolved SFR maps show NGC\,1427A moving from a bursty phase toward quenching, with the northern clump exhibiting slightly stronger H$\alpha$ suppression, as expected for an outer region that is more easily stripped and heated. Over the past $\sim\!1$~Gyr, the SFH points to a prolonged episode of moderately elevated SFR ($\sim3$-$4\times10^{-2}~M_\odot\,\mathrm{yr^{-1}}$), consistent with the migrating starburst behavior seen in other dwarfs \citep[e.g.][]{mcquinnNATURESTARBURSTSII2010a} and plausibly fostered by interactions as NGC\,1427A traveled along a filament into Fornax. As the galaxy plunges deeper into the ICM, the combination of ongoing RPS and the prior fly-by explains both the strong recent burst and the present-day drop in H$\alpha$-traced SFR, marking the onset of environmentally driven quenching as the ISM begins to be stripped and heated.

  \item \textbf{Metallicity structure: a luminous young, enriched component atop an older metal-poor body.}
  Light-weighted metallicities exceed mass-weighted values by $\sim\!0.4\!-\!0.5$~dex, implying that recent star formation dominates the light while an older, more metal-poor component dominates the mass. On the mass-metallicity plane, the galaxy remains modestly metal-poor for its mass, consistent with metal-loaded outflows in the pre-infall phase and efficient metal retention in compact star-forming complexes in the recent burst.

\end{enumerate}

The multi-phase morphology, kinematics, extinction geometry, and time-resolved star formation together suggest the following sequence (Fig.~\ref{fig:scenario-cartoon}). NGC\,1427A likely formed as a gas-rich, metal-poor dwarf and experienced elevated, spatially migrating star formation over the past $\sim$Gyr, plausibly aided by increased encounter rates as it moved from the field into a denser infall environment. An earlier galaxy--galaxy perturbation or merger, as favored by \citet{serraMeerKATFornaxSurvey2024}, may have decoupled the gas and facilitated stripping of the CGM and outer disk over the last few $\times100~{\rm Myr}$ via RPS by the Fornax ICM, producing the long, largely starless \hi\ tail. More recently ($\sim30$--$100$ Myr), a mild localized fly-by, plausibly involving FCC\,229, may have helped torque and displace the gaseous disk, triggering a localized burst in the northeast outer disk and later activity in the main body and southwest, while further decoupling collisional phases from the old stellar body.

At the present epoch (last few $\times10$~Myr), RPS is no longer acting only on the circumgalactic gas, but has reached the inner ISM. The gas disk is misaligned, partially stripped, and shifted in front of the stars; the ionized, neutral, and dusty components are decoupled from the old stellar body; and the instantaneous SFR has declined below the SFMS. These signatures mark the onset of environmentally driven quenching, as stripping, heating, and gas removal begin to directly affect the star-forming ISM. Continued interaction with the ICM is expected to exhaust and remove the remaining cold gas, leaving a quenched, gas-poor Fornax dwarf whose subsequent evolution is dominated by the collisionless components.

This reconstruction should be read as the most consistent scenario allowed by the currently available multi-phase data, not as a unique dynamical solution. As also emphasized by \citet{serraMeerKATFornaxSurvey2024}, NGC\,1427A is a complex system, and the relative timing and strength of the evolutionary drivers remain uncertain. JWST observations of the dust-bearing ISM and a direct TRGB distance to NGC\,1427A would be especially valuable for constraining its three-dimensional location, the role of cluster tides, and the progression of RPS from the CGM into the ISM.

As a spatially resolved snapshot of this transition, NGC\,1427A provides a benchmark for mixed RPS$+$tidal pathways in low-mass cluster satellites.

\begin{acknowledgements}
We thank the referee for constructive comments that improved the clarity and balance of the manuscript. JPC thanks the ISM* group at STScI for valuable discussions on ISM tracers, dust, and PAH diagnostics. This work is based on observations collected at ESO/Paranal with VLT/MUSE (programs listed in Table~\ref{tab:observations}) and makes use of archival data from \textit{HST} (MAST), \textit{AstroSat}/UVIT (ISSDC), \textit{Spitzer} (IRSA), \textit{Herschel}, MeerKAT (SARAO; MeerKAT Fornax Survey products), NGFS/DECam (CTIO/NOIRLab), and public \textit{SRG}/eROSITA survey products. We made use of \textsc{Astropy} \citep{astropy2013, astropy2018, astropy2022}, \textsc{MPDAF} \citep{baconMPDAFMUSEPython2016}, \ZAP\ \citep{sotoZAPEnhancedPCA2016}, and \textsc{pPXF}/\textsc{vorbin} \citep{cappellariAdaptiveSpatialBinning2003, cappellariImprovingFullSpectrumPPXF2017}. We acknowledge support from the Agencia Nacional de Investigación y
Desarrollo (ANID) (CATA-Basal FB210003; Beca Doctorado Nacional for RR and JPC; Proyecto Fondecyt Regular 1231345 for JC). KF acknowledges funding from the EU Horizon 2020 Marie Sk\l{}odowska-Curie program (grant No.~101103830). YOB acknowledges support from ESO Comit\'e Mixto 2024. 
\end{acknowledgements}

\bibliographystyle{aa}
\bibliography{references}

\appendix

\section{MUSE mosaic reduction details}
\label{app:MUSEdatared}

We reduced the MUSE observations of NGC\,1427A with the ESO pipeline (v3.13.8; \citealt{weilbacherDataProcessingPipeline2020}), complemented by custom steps optimized for extended, low-surface-brightness emission. The workflow is implemented in a dedicated \textsc{python} wrapper around \texttt{esorex} to ensure consistent processing across runs. Data were organized by field (F1--F4) and reduced through basic calibration, astrometric alignment, sky modeling/subtraction (field-dependent), quality control, and final coaddition onto a common WCS.

Raw frames were ingested into a custom organizer that grouped data by field (F1--F4), matched each exposure to its calibrations, and generated \textit{set-of-frames} (\textsc{sof}) files. For each exposure, we recorded pointing, UTC start time, exposure time, frame type (target or offset sky), and observing block (OB). We also computed basic observing-context diagnostics (lunar phase, separation from bright Solar System objects, background illumination) to flag exposures susceptible to strong sky gradients. None of the OBs were affected by nearby bright sources. One OB in Field~3 displayed irregular sky behavior; three of its four exposures were rejected (Sect.~\ref{sec:field2-3_skysub}).

Calibrated products were generated with the standard recipes. \texttt{muse\_scibasic} performed bias subtraction, flat-fielding, wavelength calibration, and geometric tracing, producing pixel tables for science and sky frames. We then ran an initial \texttt{muse\_scipost} pass with sky subtraction disabled (\texttt{--skymethod=none}) and without astrometric alignment to obtain preliminary cubes and \texttt{IMAGE\_FOV} whitelight maps for inspection. Several offset-sky frames and one target exposure exhibited significant drift due to tracking errors; the affected science exposure was excluded from the final mosaic.

\subsection{Astrometric alignment}
\label{sec:align}

Astrometric registration used a calibrated \textit{HST}/ACS image of NGC\,1427A (program 9689, PI: Gregg) from MAST as the absolute reference. The ACS frame was PSF-matched to MUSE with a $0\farcs7$ FWHM Gaussian kernel, flux-scaled, and resampled to the MUSE spatial grid (0\farcs2\,pix$^{-1}$). The result was written in \texttt{IMAGE\_FOV} format for direct use with \texttt{esorex}.

We ran \texttt{muse\_exp\_align} on all science exposures to measure residual offsets relative to this reference. All but two exposures aligned within $\simeq0\farcs2$; one Field~4 exposure was affected by guiding drift, and one Field~3 exposure with a large initial offset was corrected manually.

We then defined a common \texttt{OUTPUT\_WCS} covering the full four-field mosaic and the nominal wavelength range (4700--9350\,\AA). Using this \texttt{OUTPUT\_WCS} and the derived \texttt{OFFSET\_LIST}, we reran \texttt{muse\_scipost} for each exposure (still with \texttt{--skymethod=none}). Whitelight maps from these resampled cubes were compared to the PSF-matched ACS reference to verify field-to-field consistency.

\subsection{Sky subtraction}
\label{sec:skysub}

Sky subtraction required three strategies. Fields~2 and~3 have dedicated offset-sky exposures; Field~4 contains sufficient blank sky for in-field modeling; Field~1 is filled by galaxy emission and has no offset frames. The adopted procedures are summarized below.

\subsubsection{Field~4: in-field sky modeling}
\label{sec:field4_skysub}

Field~4 includes $\sim40$\,\% blank sky, so the sky background was derived from the science frames. A sky mask was built by PSF-matching and resampling an \textit{HST}/ACS detection image to the MUSE grid and thresholding to mask detectable galaxy emission. For each exposure we ran \texttt{muse\_scipost} with \texttt{--skymethod=model --skymodel\_fraction=0.70 --skymodel\_ignore=0.08 --save=cube,skymodel}. Residual sky features were mitigated with \ZAP\ \citep{sotoZAPEnhancedPCA2016}.

Field~4 has three 967\,s exposures; one was rejected due to guiding drift. The final cube shows a reliable continuum and diffuse \Halpha\ emission in the outskirts, despite a bright sky and mild large-scale gradients above $\sim$8900\,\AA.

\begin{figure*}[!htb]
    \centering
    \includegraphics[width=0.8\textwidth]{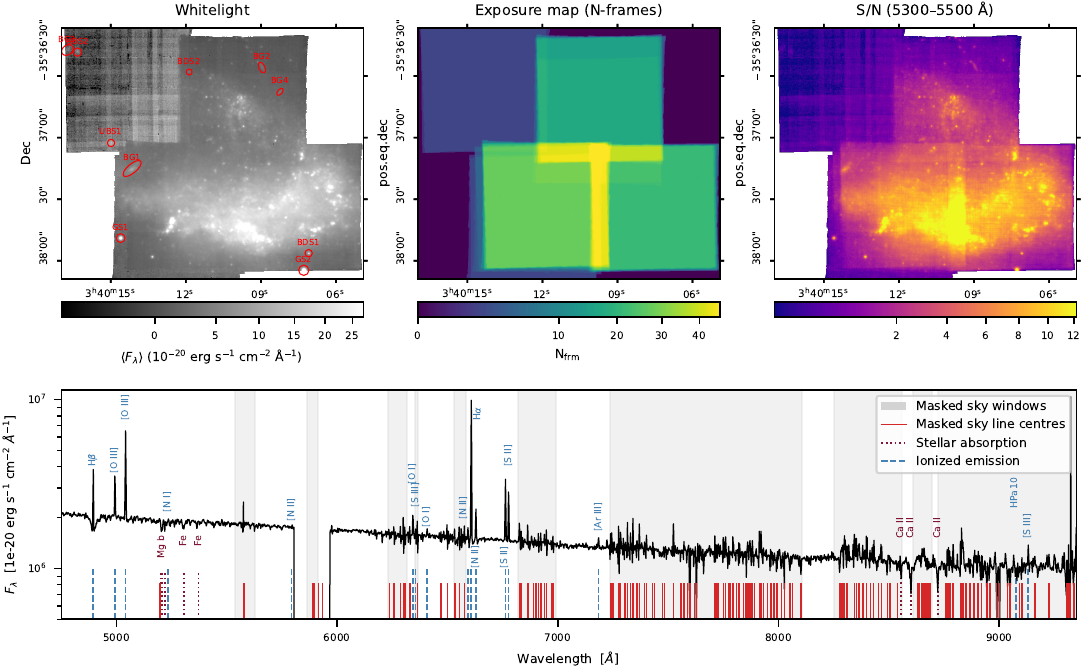}
    \caption{MUSE mosaic data-quality summary.
    \textbf{Top row (common WCS; north up, east left):} white-light image (4700--9300\,\AA; red ellipses denote masked out foreground and background sources), exposure map ($N_{\rm frm}$), and continuum S/N per spaxel (5300--5500\,\AA).
    \textbf{Bottom:} integrated spectrum over spaxels with $N_{\rm frm}>2$.
    Gray bands mark spectral windows masked due to strong sky residuals; ticks indicate masked sky lines (red), nebular lines (blue), and prominent stellar absorption features (wine).}
    \label{fig:DQ-Overview}
\end{figure*}

\subsubsection{Fields~2 and~3: offset-sky subtraction}
\label{sec:field2-3_skysub}

Fields~2 and~3 include offset-sky exposures acquired adjacent in time. Source masks for the offset frames were built from deep NGFS $u^\prime g^\prime i^\prime$ mosaics \citep{munozUNVEILINGRICHSYSTEM2015a,eigenthalerNextGenerationFornax2018a} and propagated to each offset exposure after applying the measured per-exposure shifts/drifts. For each OB we ran \texttt{muse\_create\_sky} on the masked offset exposure (\texttt{--fraction=0.95 --ignore=0.05}) to derive sky-continuum and sky-line tables; one science exposure without a paired offset used sky tables interpolated from adjacent OBs.

Per science exposure, a target mask was constructed from the PSF-matched/resampled \textit{HST} detection image (Sect.~\ref{sec:field4_skysub}) to isolate regions minimally affected by galaxy emission. Because NGC\,1427A occupies most of the FoV, we retained only the faintest $\simeq20$\,\% of spaxels as the sky mask. Science exposures were reduced with \texttt{muse\_scipost --skymethod=model --save=cube,skymodel} using the corresponding offset-derived sky tables, and residuals were removed with \ZAP\ \citep{sotoZAPEnhancedPCA2016} using an external SVD basis built from the offset cubes.

Residuals were assessed by averaging the faintest 5--10\,\% of spaxels per exposure; these should be consistent with zero within the noise. Three Field~3 frames (OB03) showed strong, rapidly varying residuals across much of the bandpass and were discarded. The behavior is inconsistent with Moon/planet proximity and is likely due to transient terrestrial light.

\subsubsection{Field~1: anchored sky model}
\label{sec:field1prepsky}

Field~1 is fully filled by galaxy emission and lacks offset frames, so we anchored its sky model to a mosaic of the reduced Fields~2 and~3 (F23), which overlap Field~1.

\paragraph{Preparatory steps.}
\textbf{(i) Velocity window:} from a coadd of minimally processed cubes (F1--F4) we measured the \Halpha\ span 1960--2120~km~s$^{-1}$.
\textbf{(ii) Trimmed \texttt{SKY\_LINES}:} we removed sky lines within $\pm80$~km~s$^{-1}$ of the galaxy-shifted H$\beta$, [O\,\textsc{iii}], \Halpha, [N\,\textsc{ii}], and [S\,\textsc{ii}] transitions to avoid fitting sky lines to the galaxy emission.
\textbf{(iii) Flux cross-check:} independent F2 and F3 coadds (sky- and \ZAP-cleaned) agree within $\lesssim9$\% over 4700--9350\,\AA\ in their overlap.
\textbf{(iv) F23 mosaic:} we built an F23 reference cube on the global \texttt{OUTPUT\_WCS}.

\paragraph{Per-exposure workflow.}
Each of the 21 Field~1 science frames was processed as follows (alignment products as in Sect.~\ref{sec:align}): (i) build a target sky mask from the PSF-matched/resampled \textit{HST} detection image, retaining the faintest $\sim20$\% of spaxels; (ii) run \texttt{muse\_scipost} with the trimmed \texttt{SKY\_LINES} and \texttt{SKY\_CONTINUUM}=0 (line-only subtraction); (iii) derive a continuum model anchored to F23 by subtracting the F23 mosaic within the overlap, selecting the faintest spaxels in the overlap$\cap$mask (typically $\sim$5\% of the field), and median-combining their spectra; (iv) rerun \texttt{muse\_scipost} with the resulting \texttt{SKY\_CONTINUUM}; (v) remove residuals with \ZAP, using an external-sky cube (\texttt{DATACUBE\_FINAL}$-$F23) to build the SVD basis.

The coadd of all Field~1 exposures matches the F2 and F3 coadds within the overlap area within $\lesssim6$\%, yielding a uniform flux scale across the final mosaic.

\subsection{Exposure combination}
\label{sec:combination}

We produced two mosaics on the common \texttt{OUTPUT\_WCS}: (i) a \ZAP-cleaned cube used for the science analysis, and (ii) a no-\ZAP\ control for validation. Individual cubes were co-added with MPDAF \citep{baconMPDAFMUSEPython2016} using \texttt{CubeList.combine} with \texttt{nmax=2}, \texttt{nclip=4.0}, \texttt{nstop=2}, and \texttt{var=propagate}. Figure~\ref{fig:DQ-Overview} summarizes the mosaic quality.

\section{DAP configuration and definitions}
\label{app:DAPmethods}

Table~\ref{tab:dap-setup} summarises the configuration referenced in Sect.~\ref{sec:data_analysis}. We report only the parameters needed for reproducibility. The \textsc{E-MILES} templates are convolved to the wavelength-dependent MUSE LSF prior to fitting following \citetalias{emsellemPHANGSMUSESurveyProbing2022}. We adopt an LMC-like total-to-selective ratio $R_V=3.41$ \citep{gordonQuantitativeComparisonSmall2003} throughout, and use the same $k(\lambda)$ prescription in the stellar- and gas-phase corrections.

Because NGC\,1427A is a dwarf galaxy, we use \textsc{E-MILES} template subsets that extend farther into the metal-poor regime than the grids adopted in \citetalias{emsellemPHANGSMUSESurveyProbing2022}. Maintaining the same logic, we use two grids, one coarse and one fine, tailored to the two main stellar-continuum steps. For the stellar-kinematics fit, we use a compact grid of 40 SSPs ($8$ ages $\times$ $5$ metallicities), with ages
$t=\{0.15,\allowbreak 0.30,\allowbreak 0.60,\allowbreak 1.00,\allowbreak 2.00,\allowbreak 3.75,\allowbreak 7.00,\allowbreak 13.5\}$~Gyr
and metallicities
$[\mathrm{Z/H}]=\{-2.27,\allowbreak -1.26,\allowbreak -0.35,\allowbreak +0.06,\allowbreak +0.26\}$.
For the stellar-population fit, we use a denser grid of 117 SSPs ($13$ ages $\times$ $9$ metallicities), with ages
$t=\{0.03,\allowbreak 0.05,\allowbreak 0.08,\allowbreak 0.15,\allowbreak 0.25,\allowbreak 0.40,\allowbreak 0.60,\allowbreak 1.0,\allowbreak 1.75,\allowbreak 3.0,\allowbreak 5.0,\allowbreak 8.5,\allowbreak 13.5\}$~Gyr
and metallicities
$[\mathrm{Z/H}]=\{-2.27,\allowbreak -1.79,\allowbreak -1.49,\allowbreak -1.26,\allowbreak -0.96,\allowbreak -0.66,\allowbreak -0.35,\allowbreak +0.06,\allowbreak +0.26\}$.
We omit the most metal-rich \textsc{E-MILES} node, $[\mathrm{Z/H}]=+0.40$, because it is flagged by the model safe-range criteria and is not expected to be relevant for the bulk of the stellar mass in NGC\,1427A. Such metallicities may still be relevant for compact, efficiently enriched stellar complexes \citep[e.g.][]{Fahrion2026a}.

\begin{table}
\caption{Core \textsc{tardis}/\textsc{pPXF} configuration adopted in this work.}
\label{tab:dap-setup}
\centering
\small
\setlength{\tabcolsep}{4pt}
\renewcommand{\arraystretch}{1.15}
\begin{tabular}{p{0.18\columnwidth}p{0.18\columnwidth}p{0.50\columnwidth}}
\hline\hline
Block & Parameter & Value \\
\hline
Global
& $v_{\rm sys}$ & $2035$~km~s$^{-1}$ \\
& Foreground $E(B{-}V)$ & $0.01$ \\
& LSF treatment & Templates convolved to the wavelength-dependent MUSE LSF \citepalias{emsellemPHANGSMUSESurveyProbing2022} \\
\hline
Binning
& Target S/N & $35$ (also $60$ and $90$ for robustness) \\
& S/N window & 5300--5500\,\AA \\
\hline
Kinematics
& Templates & 40-grid \\
& $\lambda$ range & 4800--7000\,\AA \\
& Moments & $V,\sigma,h_3,h_4$ \\
& Polynomials & additive / multiplicative $=12/0$ (as \citetalias{emsellemPHANGSMUSESurveyProbing2022}) \\
\hline
Attenuation
& Templates & 40-grid \\
& $\lambda$ range & 4800--7000\,\AA \\
& $R_V$ & 3.41 (LMC-like) \\
& Polynomials & additive / multiplicative $=0/0$ \\
\hline
Populations
& Templates & 117-grid (as \citetalias{emsellemPHANGSMUSESurveyProbing2022}) \\
& $\lambda$ range & 4760--9350\,\AA\ (masked windows; Fig.~\ref{fig:DQ-Overview}) \\
& Regularisation & none \\
& MC realizations & 20 \\
& Polynomials & additive / multiplicative $=0/12$ (as \citetalias{emsellemPHANGSMUSESurveyProbing2022}) \\
\hline
Emission lines
& $\lambda$ range & 4760--9350\,\AA; brightest observed-frame sky lines masked using the UVES atlas \citep{hanuschikUVESSkyLineAtlas2003} \\
& Kinematic tying & 3 families (H recombination, low-ionization, high-ionization; cf. \citealt{belfioreDataAnalysisPipeline2019}) \\
& Detection & robust if not fully compromised by masking and S/N$>5$ in at least one bin, with S/N$\equiv F_{\rm line}/\sigma(F_{\rm line})$ \citepalias{emsellemPHANGSMUSESurveyProbing2022} \\
\hline
Derived
& $E(B{-}V)_{\rm gas}$ & Balmer decrement (case B) with a Calzetti-type curve \citepalias[][]{calzettiDustContentOpacity2000} evaluated at $R_V=3.41$ \\
& SFR & \Halpha\ calibration from \citet{kennicuttStarFormationMilky2012} \\
\hline
\end{tabular}
\tablefoot{The fitting sequence follows \citetalias{emsellemPHANGSMUSESurveyProbing2022}, but the stellar-template grids are extended to lower metallicity. The main masked regions are indicated in Fig.~\ref{fig:DQ-Overview}.}
\end{table}

\subsection{Nebular reddening and \Halpha-based SFRs}
\label{app:reddening_sfr}

Nebular reddening is estimated from the Balmer decrement under case-B recombination, adopting $(\mathrm{H}\alpha/\mathrm{H}\beta)_{\rm int}=2.86$ and a Calzetti-type attenuation curve \citepalias[][]{calzettiDustContentOpacity2000} evaluated with our adopted $R_V=3.41$:
\begin{equation}
E(B{-}V)_{\rm gas} =
\frac{2.5}{k(\lambda_{\mathrm{H}\beta})-k(\lambda_{\mathrm{H}\alpha})}
\log_{10}\!\left[\frac{(\mathrm{H}\alpha/\mathrm{H}\beta)_{\rm obs}}
{(\mathrm{H}\alpha/\mathrm{H}\beta)_{\rm int}}\right].
\end{equation}
The extinction-corrected \Halpha\ luminosity is
\[
L_{{\rm H}\alpha}=4\pi D^2\,F_{{\rm H}\alpha,{\rm obs}}\;10^{0.4\,k(\lambda_{\rm H\alpha})\,E(B{-}V)_{\rm gas}},
\]
where $D$ is the adopted distance to NGC\,1427A ($D=17\pm2.5$\,Mpc). We convert to SFR using \citet{kennicuttStarFormationMilky2012} (see also \citealt{chomiukUnificationStarFormation2011}):
\[
\mathrm{SFR}_{{\rm H}\alpha}\,[M_\odot\,\mathrm{yr}^{-1}] = (5.5\times10^{-42})\,L_{{\rm H}\alpha}\,[\mathrm{erg\,s^{-1}}].
\]
Uncertainties propagate the line-flux errors and the reddening term; maps are additionally masked where either \Halpha\ or H$\beta$ fails the detection criterion above.

\subsection{Stellar-population summary maps and region-integrated PDFs, SFHs, and MDFs}
\label{app:pdfreg}

\paragraph{Per-bin stellar-population summary maps.}
We construct the stellar-population maps from the non-negative pPXF template weights, averaged over the Monte Carlo (MC) ensemble, similar to \citetalias{emsellemPHANGSMUSESurveyProbing2022}. Let $i$ index Voronoi bins, $j$ index SSP templates on the native library grid, and $r$ index MC realizations. For each $(i,r)$, the fit returns a weight $w_{irj}$ for template $j$. Using the template-intrinsic present-day stellar mass $M^{\ast}_{j}$, logarithmic age
\[
\mathcal{A}_{j}\equiv \log_{10}(\mathrm{age}_{j}/\mathrm{yr}),
\]
and metallicity
\[
\mathcal{Z}_{j}\equiv [\mathrm{Z/H}]_{j},
\]
we define the present-day stellar mass contributed by template $j$ in bin $i$ and realization $r$ as
\[
m_{irj} = w_{irj}\, M^{\ast}_{j}.
\]
We then average over MC realizations,
\[
\bar m_{ij} = \frac{1}{N_{\mathrm{MC}}}\sum_{r=1}^{N_{\mathrm{MC}}} m_{irj}.
\]

Let $A_i$ be the projected area of Voronoi bin $i$ in $\mathrm{kpc}^2$. The stellar mass surface density map is
\[
\Sigma_{\ast,i} = \frac{1}{A_i}\sum_j \bar m_{ij}.
\]
The mass-weighted mean logarithmic age and metallicity maps are
\[
\langle \mathcal{A} \rangle_{M,i} =
\frac{\sum_j \bar m_{ij}\,\mathcal{A}_j}{\sum_j \bar m_{ij}},
\qquad
\langle \mathcal{Z} \rangle_{M,i} =
\frac{\sum_j \bar m_{ij}\,\mathcal{Z}_j}{\sum_j \bar m_{ij}}.
\]

For light-weighted quantities, we convert each template mass into light using the template mass-to-light ratio $\Upsilon_j \equiv M^{\ast}_{j}/L_{V,j}$,
\[
\bar \ell_{ij} =
\frac{1}{N_{\mathrm{MC}}}\sum_{r=1}^{N_{\mathrm{MC}}}\frac{m_{irj}}{\Upsilon_j},
\]
and replace $\bar m_{ij}$ by $\bar \ell_{ij}$ in the expressions above. All map-level stellar-population summaries are computed from the MC-averaged quantities rather than from a single best-fit solution.

These per-bin quantities are used both to construct the spatial maps discussed in Sect.~3.3 and to derive region-integrated stellar-population summaries. Using the MC-averaged present-day stellar mass contributions $\bar m_{ij}$ defined above, we combine the contributions from all Voronoi bins within a selected region and normalize them to the SSP library grid in age and metallicity. For a given region, defined by a binary selection mask $b_i$, the region-integrated template mass is
\begin{equation}
\bar m^{\rm reg}_j = \sum_i b_i\,\bar m_{ij},
\end{equation}
and the region-integrated, mass-weighted PDF is
\begin{equation}
\mathrm{PDF}^{\rm reg}_j \equiv \frac{\bar m^{\rm reg}_j}{\sum_k \bar m^{\rm reg}_k}.
\end{equation}
By construction, $\sum_j \mathrm{PDF}^{\rm reg}_j=1$, so $\mathrm{PDF}^{\rm reg}_j$ represents the fraction of the region's present-day stellar mass assigned to SSP template $j$.

The metallicity distribution function (MDF) is obtained by marginalizing $\mathrm{PDF}^{\rm reg}$ over age and scaling by the region stellar mass; it is therefore expressed in present-day stellar mass. The star-formation history (SFH) is obtained by (i) converting present-day stellar mass in each template to an initially formed mass using the SSP library living-mass fraction, (ii) marginalizing over metallicity, and (iii) dividing by the linear time-bin width to yield $\mathrm{SFR}(t)$.

\section{Ancillary data products and calibrations}
\label{app:ancillary}

\subsection{AstroSat/UVIT far-UV imaging}
\label{app:uvit}

NGC\,1427A was observed with AstroSat/UVIT in the F148W filter (mean wavelength $\lambda_{\rm mean}=1481$~\AA, bandwidth $\Delta\lambda\simeq500$~\AA) in two epochs: 30~Nov~2022 (26883.80~s) and 28~Jan~2023 (37004.35~s), for a total exposure time of 63888.15~s. Level-1 data were corrected for drift, flat-field, and distortion \citep[cf.][]{postma2017_CCDLAB}, and combined into the final science mosaic; 824.44~s from 28~Jan~2023 were discarded due to large distortions. The final image has pixel scale $0.4167\arcsec\,{\rm pix}^{-1}$ and PSF FWHM $\simeq1.5\arcsec$, and the astrometry was refined against \textit{Gaia}~DR3 to $\simeq0.4\arcsec$ accuracy. Photometric calibration follows the mission prescriptions for F148W \citep{tandonInorbitCalibrationsUltraviolet2017, tandonAdditionalCalibrationUltraviolet2020}, reporting AB magnitudes.

We measure integrated UVIT photometry in three apertures defined on the NGFS optical imaging: (i) a circular aperture of radius $15\arcsec$ centered on the ``northern clump'', (ii) a ``main body'' aperture defined by the $\mu=\mu_0+2.0$ isophote, and (iii) a ``full galaxy'' aperture out to $\mu=\mu_0+4.5$ to include low-surface-brightness outskirts while mitigating contamination from neighboring sources. The background is measured in an annulus between $140\arcsec$ and $200\arcsec$.

We convert FUV luminosity densities to SFRs using a continuous-star-formation calibration appropriate for $\gtrsim100$~Myr timescales (\citealt{kennicuttStarFormationMilky2012}; cf. \citealt{rampazzoDoradoItsMember2022}):
\begin{equation}
\label{eq:FUV_SFR}
{\rm SFR}_{\rm FUV} = K_\nu\,L_\nu({\rm FUV}),
\end{equation}
with $K_\nu=9.04\times10^{-29}\ M_\odot\,{\rm yr^{-1}}\ ({\rm erg\ s^{-1}\ Hz^{-1}})^{-1}$. We obtain $L_\nu$ from the extinction-corrected $F_\lambda$ evaluated at $\lambda_{\rm mean}$ using the standard $F_\nu=(\lambda^2/c)\,F_\lambda$ relation and $L_\nu=4\pi D^2F_\nu$.

For attenuation corrections, we apply a Milky Way foreground reddening of $E(B{-}V)_{\rm MW}=0.01$ \citep{schlaflyMeasuringReddeningSloan2011} and an internal color excess tied to the nebular Balmer-decrement estimate (Sec.~\ref{sec:data_analysis}; App.~\ref{app:reddening_sfr}; Fig.~\ref{fig:dust-properties}). For integrated values we adopt a uniform $E(B{-}V)_{\rm int}=0.04\pm0.02$ representative of the spatially averaged nebular reddening. Bandpass-weighted attenuation coefficients are computed across the F148W transmission using the \citepalias[][]{cardelliRelationshipInfraredOptical1989} relation with $R_V=3.1$ for the Milky Way term and a Calzetti-type curve with $R_V=3.41$ \citepalias[][]{calzettiDustContentOpacity2000} for the internal term, yielding $\langle k_{\rm FUV}^{\rm MW}\rangle=9.05$ and $\langle k_{\rm FUV}\rangle=10.03$ (mag per unit $E(B{-}V)$).

The observed aperture-integrated FUV photometry is $m_{\rm AB}=18.319\pm0.011$ for the northern clump ($r=15\arcsec$), $m_{\rm AB}=15.696\pm0.010$ for the main body ($\mu=\mu_0+2.0$), and $m_{\rm AB}=15.457\pm0.010$ for the full galaxy aperture ($\mu=\mu_0+4.5$). After applying the attenuation corrections above and using Eq.~\ref{eq:FUV_SFR}, we obtain ${\rm SFR}_{\rm FUV}=(8\pm3)\times10^{-3}\ M_\odot\,{\rm yr^{-1}}$ for the northern clump, ${\rm SFR}_{\rm FUV}=(94\pm3.3)\times10^{-3}\ M_\odot\,{\rm yr^{-1}}$ for the main body, and ${\rm SFR}_{\rm FUV}=(117\pm0.41)\times10^{-3}\ M_\odot\,{\rm yr^{-1}}$ for the full galaxy aperture. The quoted uncertainties propagate the photometric calibration, the adopted attenuation terms (Milky Way plus internal), and the distance uncertainty.

\subsection{Optical imaging}
\label{app:optical}

\paragraph{HST imaging.}
NGC\,1427A has public \textit{HST} imaging with ACS in broad optical filters (F475W, F625W, F775W, F850LP) and the narrow-band F660N (\Halpha), as well as WFPC2/F336W in the near-UV. In this work, we use these data as an ancillary high-resolution reference to refine the astrometric alignment of the MUSE mosaic and to verify the location and compactness of bright \hii\ regions (Sect.~\ref{sec:data_muse}; App.~\ref{app:MUSEdatared}). We rely on the calibrated level-3 products available via MAST and refer the reader to the instrument documentation and to \citet{moraStarburstCoreGalaxy2015} for analyses of the young stellar complexes.

\begin{figure}
    \centering
    \includegraphics[width=\linewidth]{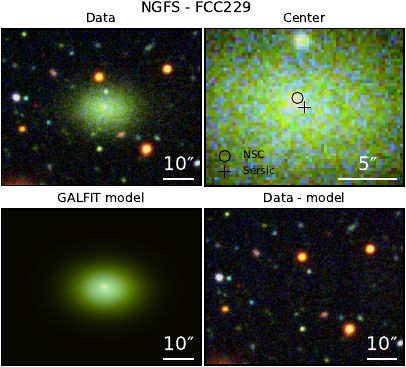}
    \caption{
    \textit{Top:} NGFS $u^\prime g^\prime i^\prime$ composite of FCC\,229 and a zoomed-in view of the nucleus, highlighting the offset between the nuclear star cluster (circle) and the S\'ersic center (cross).
    \textit{Bottom:} \textsc{GALFIT} model and residuals for the same region.
    }
    \label{fig:FCC229_morph}
\end{figure}

\paragraph{NGFS wide-field imaging and neighboring dwarfs.}
Wide-field optical context is provided by NGFS $u^\prime g^\prime i^\prime$ imaging \citep{munozUNVEILINGRICHSYSTEM2015a, eigenthalerNextGenerationFornax2018a}. The NGFS data reveal several very low-surface-brightness dwarfs in projection near NGC\,1427A (Fig.~\ref{fig:ngc1427a-ngfs-isophotes}); in the main text, we restrict ourselves to their qualitative use as potential perturbers in projection. Among these systems, FCC\,229 is of particular interest because it hosts a compact nucleus whose centroid appears offset from the galaxy light distribution (Fig.~\ref{fig:FCC229_morph}).

\paragraph{Structural fit of FCC\,229.}
To quantify the apparent nuclear offset, we fit FCC\,229 with \textsc{GALFIT} \citep{pengDetailedStructuralDecomposition2002, pengDetailedDecompositionGalaxy2010} on the NGFS $u^\prime g^\prime i^\prime$ images. In $g^\prime$ and $i^\prime$ we model the galaxy with a S\'ersic component plus a central unresolved component (PSF) to represent the nuclear star cluster, while in $u^\prime$ we adopt a single S\'ersic model given the lower S/N and weaker nucleus contrast. The fitted NSC has $m_{i^\prime}=23.5$~mag and is displaced by $\simeq1\arcsec$ from the best-fitting S\'ersic center (Fig.~\ref{fig:FCC229_morph}). This offset is significant compared to the $\simeq1.1$--$1.3\arcsec$ seeing in the $g^\prime$ and $i^\prime$ images. Within the uncertainties, the NSC color is consistent with the integrated galaxy light. We use this offset only as a qualitative consistency check; nuclear offsets of this scale are not uncommon in nucleated dwarfs \citep[e.g.][]{binggeliOffcenterNucleiDwarf2000} and are not by themselves diagnostic of an interaction. The implications are discussed in Sect.~\ref{sec:discussion}.

\subsection{Infrared data: Spitzer and Herschel}
\label{app:ir}

We use archival \textit{Spitzer} IRAC imaging (3.6, 4.5, and 8.0~$\mu$m) and \textit{Spitzer} IRS spectral mapping, together with \textit{Herschel} PACS and SPIRE maps and aperture photometry from HeFoCS \citep{sivanandamTracingRampressureStripping2014, fullerHerschelFornaxCluster2014}. We do not re-reduce the basic calibrated data; our processing is limited to registration, PSF matching, aperture definition, and spectral extraction for the IRS cubes.

\paragraph{IRAC imaging (PAHs and stellar continuum).}
To isolate the dust-dominated 8~$\mu$m component, we subtract the stellar continuum contribution using the prescription of \citet{helouAnatomyStarFormation2004}, using the 3.6~$\mu$m image. Given the warm dust temperatures inferred from the IRS-based analysis below, the warm dust continuum falls steeply toward wavelengths $\lesssim10~\mu$m, so the continuum-subtracted 8~$\mu$m image primarily traces PAH-bearing dust with little contamination from the warm dust continuum. For multi-wavelength comparisons, the IRAC 8~$\mu$m, UVIT FUV, and MUSE \Halpha\ images are registered to the MUSE astrometric frame and PSF-matched to a common $2\arcsec$ FWHM Gaussian, set by the IRAC 8~$\mu$m resolution (see Fig.~\ref{fig:FUV-HA-8UM}).

\begin{figure}
    \centering
    \includegraphics[width=\linewidth]{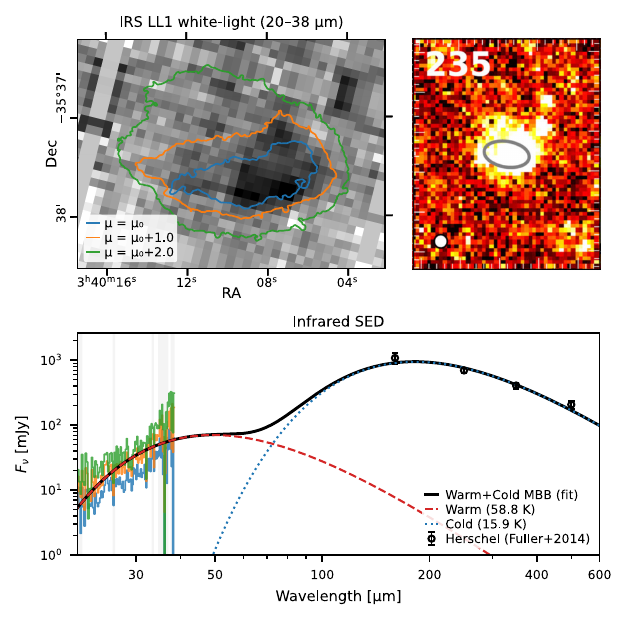}
    \caption{Mid- to far-IR morphology and SED of NGC\,1427A. Top left: IRS LL1 white-light map, tracing warm dust; optical isophotes at $\mu=\mu_0$, $\mu_0{+}1$, and $\mu_0{+}2$~mag~arcsec$^{-2}$ are overlaid. The warm dust is extended and asymmetric, with emission in the main body, the northern clump and the bridge region, and an enhancement toward the southwest. Top right: Herschel/SPIRE 250~$\mu$m image from HeFoCS \citep{fullerHerschelFornaxCluster2014}; the photometric aperture used in that work is shown and is comparable in scale to our $\mu_0{+}1$ isophote. Bottom: IR SED combining IRS spectra extracted within the three optical isophotes (colors as in the top-left panel) and the integrated Herschel photometry (black circles). A two-component modified blackbody fit with $\beta=2$ yields $T_{\mathrm{warm}}=59\pm3$~K and $T_{\mathrm{cold}}=15.9\pm2.6$~K (masses in App.~\ref{app:ir}).}
    \label{fig:MIR-FIR-dust}
\end{figure}

\paragraph{IRS mapping and warm-dust morphology.}
We retrieved the IRS pipeline products from IRSA and built spectral cubes using CUBISM \citep{smithSpectralMappingReconstruction2007}, including background subtraction, bad pixel masking, and cube reconstruction. From the IRS long-low (19.5--38~$\mu$m) mapping we construct a white-light image to trace warm dust and compare it to optical isophotes and to the MUSE nebular emission. To build an IR SED, we combine IRS spectra extracted within optical apertures ($\mu=\mu_0$, $\mu_0+1$, and $\mu_0+2$) with the integrated \textit{Herschel} photometry (see Fig.~\ref{fig:MIR-FIR-dust}). Emission-line regions in the IRS spectra are masked for continuum-based SED fitting. For a full analysis of the MIR lines, refer to \citet{sivanandamTracingRampressureStripping2014}.

\paragraph{Herschel photometry and two-component modified-blackbody fit.}
We adopt the aperture photometry reported by \citet{fullerHerschelFornaxCluster2014}; their optical-sized aperture is comparable in scale to our $\mu=\mu_0+1$ isophote. Combining those fluxes with the IRS long-low continuum yields a global mid- to far-IR SED. We fit the SED with a two-component modified blackbody with fixed emissivity index $\beta=2$, obtaining characteristic dust temperatures of $T_{\rm cold}=15.9\pm2.6$~K and $T_{\rm warm}=59\pm3$~K, and corresponding dust masses $M_{{\rm d,cold}}=(4.0\pm2.5)\times10^6~M_\odot$ and $M_{{\rm d,warm}}=(4.2\pm2.2)\times10^2~M_\odot$. The quoted uncertainties include the fit covariance and the distance uncertainty. The $\mu_0+2$ IRS extraction contains roughly twice the 20--38~$\mu$m flux of the $\mu_0+1$ extraction, implying that optical-sized apertures can underestimate the total dust emission; consequently, the total dust mass may plausibly be larger by up to a factor of $\sim2$ if the cold component is similarly extended.

\subsection{Radio and millimeter observations}
\label{app:radio_mm}

\paragraph{Neutral atomic gas: MeerKAT \hi.}
We use \hi\ data products from the MeerKAT Fornax Survey \citep{serraMeerKATFornaxSurvey2023a, kleinerMeerKATFornaxSurvey2023a} and the dedicated analysis presented by \citet{serraMeerKATFornaxSurvey2024}. These observations trace the 21\,cm line with high surface-brightness sensitivity and an angular resolution of $\sim6\arcsec$ (corresponding to $\sim0.5$\,kpc at our adopted distance). In this work, we make use of the publicly released \hi products (total intensity and velocity-field maps) to provide the large-scale neutral-gas morphology and kinematic context for comparison with the MUSE stellar and ionized-gas diagnostics (Sect.~\ref{sec:discussion}). We refer the reader to the above references for observational setup and data reduction details.

\paragraph{Molecular gas: ALMA CO upper limits.}
Constraints on the cold molecular phase are taken from ALMA CO(1--0) observations reported by \citet{zabelALMAFornaxCluster2019a} as part of the ALMA Fornax Cluster Survey. NGC\,1427A was observed in Band~3 at 115\,GHz, with a primary beam of $\sim55\arcsec$ FWHM centered on the galaxy, covering the central stellar body. No CO(1--0) emission is detected. For non-detections, \citet{zabelALMAFornaxCluster2019a} assume a Gaussian line profile with FWHM $50~\mathrm{km\,s^{-1}}$ to derive 3$\sigma$ upper limits on the integrated CO flux.

The CO limits are converted to H$_2$ masses using a metallicity-dependent CO--to--H$_2$ conversion factor following \citet{accursoDerivingMultivariateACO2017}, combined with a stellar-mass-based metallicity estimate and an assumed offset from the star-forming main sequence of $\Delta_{\rm MS}=0$. These assumptions are consistent with the stellar population and star-formation properties derived in this work (Sects.~\ref{sec:stellar_pop_properties} and \ref{sec:sfr_results}). For NGC\,1427A, \citet{zabelALMAFornaxCluster2019a} report $\log_{10}(M_{\rm H_2}/M_\odot) < 7.42$ at 19.95\,Mpc, which rescales to $\log_{10}(M_{\rm H_2}/M_\odot) < 7.28$ at the distance adopted here. The dependence of this constraint on the adopted $\alpha_{\rm CO}$ in the low-metallicity regime is revisited in Sect.~\ref{sec:rps-gas-budget}.

A faint 3\,mm continuum source is detected at the optical position of the galaxy; given the available data, this emission could be associated with compact star formation (cf. the \Halpha\ and 8~$\mu$m emission in Fig.~\ref{fig:FUV-HA-8UM}) or with a background source.

\section{Kinematic position-angle estimates from velocity maps}
\label{app:kinpa}

To characterize the orientation of the stellar, ionized-gas, and neutral-gas kinematics in a homogeneous way, we estimate the dominant large-scale velocity gradient directly from the projected velocity maps. This is intended as a simple first-order characterization of the kinematic orientation, not as a substitute for a full kinematic model.

\begin{figure}[!htb]
    \centering
    \includegraphics[width=\linewidth]{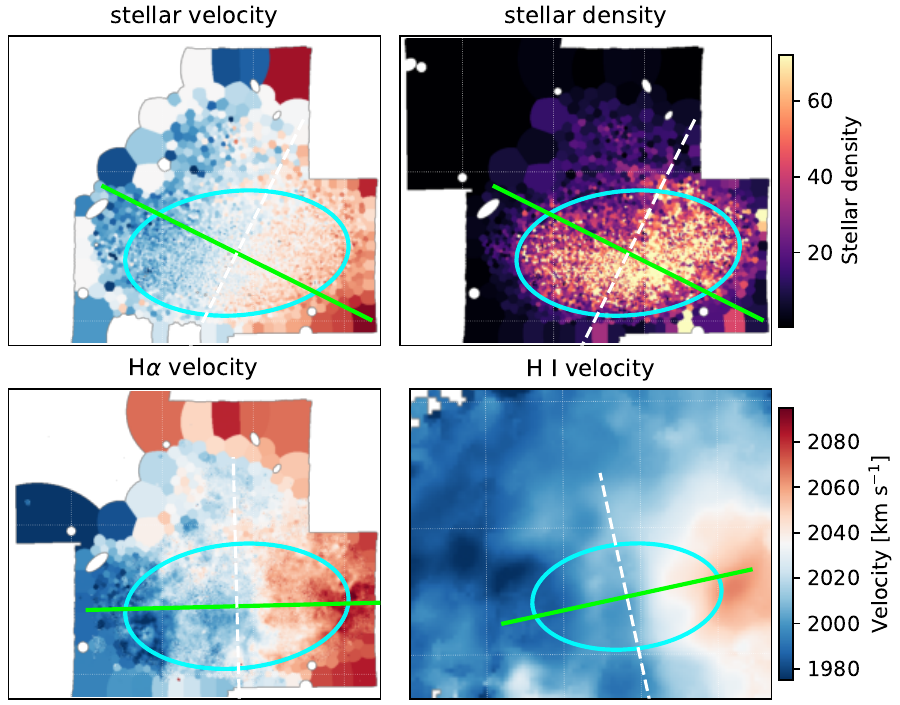}
    \caption{Velocity-based estimate of the kinematic position angle of NGC\,1427A. The panels show the stellar velocity field (top left), stellar surface-density map (top right), H$\alpha$ velocity field (bottom left), and \hi\ velocity field (bottom right), all displayed over the same sky region. The cyan ellipse marks the common aperture adopted for the fit. The green line indicates the fitted kinematic major axis, and the white dashed line shows the axis of rotation.}
    \label{fig:kin_pa_overview}
\end{figure}

We define an ad hoc elliptical aperture (see Fig.~\ref{fig:kin_pa_overview}) enclosing the main stellar body of NGC\,1427A, guided by the stellar surface-density map and applied identically to the stellar, H$\alpha$, and \hi\ velocity fields. Within this aperture, we estimate the projected velocity gradient by fitting a 2D least-squares plane in local sky coordinates and define the kinematic major-axis position angle, ${\rm PA}_{\rm kin}$, as the direction of that gradient, measured east of north. The corresponding rotation axis is perpendicular.

For all three tracers, we derive a formal covariance-based uncertainty, $\sigma_{\rm cov}$, from the fitted plane. For the stellar and H$\alpha$ velocity maps, we also propagate the available per-pixel velocity uncertainties via Monte Carlo realizations. In each realization, the velocities are perturbed according to their formal uncertainties, and the kinematic position angle is remeasured. We adopt $N=500$ realizations and define $\sigma_{\rm MC}$ from the resulting distribution of ${\rm PA}_{\rm kin}$. For these two tracers, we report a total uncertainty
\begin{equation*}
\sigma_{\rm tot} = \sqrt{\sigma_{\rm cov}^2 + \sigma_{\rm MC}^2}.
\end{equation*}
For the \hi\ moment-1 map, no equivalent per-pixel velocity-error map is available in the present data products, so we report only the covariance-based uncertainty. For a more rigorous treatment of the \hi\ kinematics, see \citet{serraMeerKATFornaxSurvey2024}. Our result is in agreement.

Applying this procedure within the adopted aperture yields kinematic major-axis position angles of $63.6^\circ \pm 1.4^\circ$ for the stellar component, $91.4^\circ \pm 0.5^\circ$ for H$\alpha$, and $102.0^\circ \pm 1.8^\circ$ for \hi. The corresponding rotation axes are therefore at $-26.4^\circ$, $1.4^\circ$, and $12.0^\circ$, respectively. These values confirm that the stellar kinematics are significantly misaligned with respect to both gas phases (by $\sim30^\circ$), while the H$\alpha$ and \hi\ kinematics remain more nearly aligned with each other. As shown in the zero-velocity curves from Fig.~\ref{fig:ngc1427a_15panel}, there is warping in the ${\rm PA}_{\rm kin}$ of all distributions, most notably for \hi. This is discussed in detail in \citet{serraMeerKATFornaxSurvey2024}.

\end{document}